\documentclass[11pt]{article}

\usepackage{jheppub}
\usepackage{slashed,dsfont,bm}
\usepackage{mathtools}
\edef\restoreparindent{\parindent=\the\parindent\relax}
\usepackage{parskip}
\restoreparindent
\DeclarePairedDelimiter{\ceil}{\lceil}{\rceil}
\DeclarePairedDelimiter{\floor}{\lfloor}{\rfloor}

\usepackage[whole]{bxcjkjatype} 

\usepackage[usenames,dvipsnames,svgnames]{xcolor}

\usepackage{tabularx}
\newcolumntype{Y}{>{\centering\arraybackslash}X}

\usepackage{booktabs,multirow,colortbl}

\usepackage{tikz}
\usetikzlibrary{decorations,decorations.pathmorphing}
\tikzset{snake it/.style={decorate, decoration=snake}}

\def\i{\textrm{i}}
\newcommand{\dd}{\mathrm{d}}

\def\BH{\mathbb{H}}

\def\BS{\mathbb{S}}
\def\CM{\mathcal{M}}

\title{Free energy and defect $C$-theorem in free fermion}

\author[]{Yoshiki Sato}

\affiliation[]{Physics Division, National Center for Theoretical Sciences, National Tsing-Hua University, \\
Hsinchu 30013, Taiwan}

\abstract{We describe a $p$-dimensional conformal defect of a free Dirac fermion on a $d$-dimensional flat space as boundary conditions on a conformally equivalent space $\mathbb{H}^{p+1} \times \mathbb{S}^{d-p-1}$.
We classify allowed boundary conditions and find that the Dirichlet type of boundary conditions always exists while the Neumann type of boundary condition exists only for a two-codimensional defect.
For the two-codimensional defect, a double trace deformation triggers a renormalization group flow from the Neumann boundary condition to the Dirichlet boundary condition, and the free energy at UV fixed point is always larger than that at IR fixed point.
This provides us with further support of a conjectured $C$-theorem in DCFT.
}

\begin{document}
\maketitle

\section{Introduction}

A renormalization group (RG) is a fundamental concept in quantum field theory (QFT).
By a deformation by a relevant operator $\mathcal{O}$, 
\begin{align}
    I_\text{CFT} + \lambda  \int \! \dd^d x \, \mathcal{O} \,,
\end{align}
a conformal field theory (CFT) at a UV fixed point flows to a CFT at an IR fixed point.
The RG flow is not reversible, and this irreversibility implies the existence of a monotonic function, known as a $C$-function.
A $C$-theorem which states the existence of the $C$-function is pioneered by Zamolodchikov \cite{Zamolodchikov:1986gt} in two-dimensional QFTs and is extended to higher dimensions \cite{Cardy:1988cwa,Komargodski:2011vj,Jafferis:2011zi,Klebanov:2011gs,Myers:2010xs,Myers:2010tj}.
Combining $a$-theorem \cite{Cardy:1988cwa,Komargodski:2011vj} which is a $C$-theorem in $d=4$ and  
$F$-theorem \cite{Jafferis:2011zi,Klebanov:2011gs} which is a $C$-theorem in $d=3$, a generalised $F$-theorem is conjectured \cite{Giombi:2014xxa}:
A free energy on a sphere $\mathbb{S}^d$, 
\begin{align}
    \tilde{F} = \sin \left( \frac{\pi d}{2}\right) \log Z [\mathbb{S}^d] \,,
\end{align}
is a $C$-function and satisfies the monotonic relation
\begin{align}
    \tilde{F}_{\text{UV}} \geq \tilde{F}_{\text{IR}} \,.
\end{align}
The generalised $F$-theorem states monotonicity of an anomaly coefficient in the free energy for even $d$, while monotonicity of a finite part in the free energy for odd $d$.
The $C$-theorem can also be proved by using an information theoretical method for $d \leq 4$ \cite{Casini:2004bw,Casini:2012ei,Casini:2017vbe}.

For QFTs with $p$-dimensional defect, it is possible to consider an RG flow triggered by a relevant operator localising on the defect,
\begin{align}
    I_\text{DCFT} + \hat{\lambda}  \int \! \dd^p x \, \hat {\mathcal{O}} \,.
\end{align}
In \cite{Kobayashi:2018lil}, we conjectured that the defect free energy on a sphere, which is an increment of the free energy due to the defect, 
\begin{align} \label{defect_FE}
    \log \langle \mathcal{D}^{(p)} \rangle  = \log Z^{\text{DCFT}} [\mathbb{S}^d] - \log Z^{\text{CFT}} [\mathbb{S}^d] \,,
\end{align}
decreases under the defect RG flow.
More precisely, the universal part of the defect free energy,
\begin{align}
    \tilde{D} = \sin \left( \frac{\pi p}{2}  \right) \log |\langle \mathcal{D}^{(p)} \rangle|
\end{align}
is a $C$-function, and it decreases under the defect RG flow,
\begin{align}
    \tilde{D}_{\text{UV}} \geq \tilde{D}_{\text{IR}} \,.
\end{align}
For BCFTs with $p=d-1$, a slight modification is needed since BCFTs is defined on a hemisphere $\mathbb{HS}^d$ (a half space of the original CFTs), and the boundary free energy is introduced as\footnote{It is pointed out that the boundary free energy does not decrease monotonically under the bulk RG flow  \cite{Green:2007wr,Sato:2020upl}.}
\begin{align} \label{bounday_FE}
    \log \langle \mathcal{D}^{(p)} \rangle  = \log Z^{\text{BCFT}} [\mathbb{HS}^d] - \frac{1}{2} \log Z^{\text{CFT}} [\mathbb{S}^d] \,.
\end{align}
For BCFT, our conjecture reproduces a proved $C$-theorem in BCFT$_2$, known as $g$-theorem \cite{Affleck:1991tk,Friedan:2003yc,Casini:2016fgb}, 
and a $C$-theorem in higher dimensional BCFT \cite{Jensen:2015swa,Nozaki:2012qd,Gaiotto:2014gha,Wang:2021mdq}.
In particular, the $C$-theorem in BCFT$_3$ is proved in \cite{Jensen:2015swa} by extending \cite{Komargodski:2011vj} and in \cite{Casini:2018nym} by extending \cite{Casini:2016fgb}.
In the context of holography, a $C$-theorem in BCFT is investigated in \cite{Yamaguchi:2002pa,Takayanagi:2011zk,Fujita:2011fp,Estes:2014hka,Miao:2017gyt}.\footnote{In \cite{Estes:2014hka}, it is argued that the defect entropy, which is an increment of the entanglement entropy of the spherical region where the defect sits at the centre due to the presence of the defect, is a candidate of a $C$-function in DCFT.
The defect entropy is not a $C$-function for $p<d-1$ as pointed out \cite{Kobayashi:2018lil}. See also related works \cite{Kumar:2016jxy,Kumar:2017vjv}.
}
For DCFT with $p<d-1$, our conjecture however has passed only several checks in field theory \cite{Jensen:2015swa,Beccaria:2017rbe,Kobayashi:2018lil,Jensen:2018rxu,Wang:2020xkc,Wang:2021mdq} and holography \cite{Kumar:2016jxy,Kumar:2017vjv,Rodgers:2018mvq}.

Recently, we provide further evidence of our conjecture in a simple model, a conformally coupled scalar field \cite{Nishioka:2021uef} (See also related works \cite{Rodriguez-Gomez:2017kxf,Rodriguez-Gomez:2017aca,Gustavsson:2019zwm}).
Instead of putting the conformally coupled scalar field on the sphere, we put it on $\mathbb{H}^{p+1} \times \mathbb{S}^{q-1}$, where $q=d-p$ is a codimension of the defect, and impose boundary conditions at the boundary of $\mathbb{H}^{p+1}$.
The idea of mapping BCFTs on a flat space $\mathbb{R}^d$ to $\mathbb{H}^d$ has appeared in \cite{Herzog:2019bom,Giombi:2020rmc}, and the similar idea for DCFTs has appeared in \cite{Kapustin:2005py,Chester:2015wao,Rodriguez-Gomez:2017kxf}.
It enables us to classify allowed boundary conditions which are consistent with a recent classification \cite{Lauria:2020emq}.
The Dirichlet type boundary condition is always allowed while the Neumann type boundary condition is allowed only in  $q=1,2,3,4$.
An RG flow from the Neumann boundary condition to the Dirichlet boundary condition realised by a double trace deformation as is familiar with the AdS/CFT setup \cite{Witten:2001ua,Berkooz:2002ug,Gubser:2002zh,Gubser:2002vv,Hartman:2006dy,Diaz:2007an,Giombi:2013yva}, and the defect free energy with the Neumann boundary condition is always larger than that of the Dirichlet boundary condition. 

In this paper, we extend our analysis \cite{Nishioka:2021uef} to a free fermion to provide a further check of our conjecture.
To compare with a scalar field, a defect free energy of a free fermion in higher dimensions has been studied only in BCFT \cite{Rodriguez-Gomez:2017aca,Herzog:2019bom} and DCFT with a two-codimensional defect in the context of entanglement entropy \cite{Klebanov:2011uf,Beccaria:2017dmw}.
In particular, the existence of a nontrivial boundary condition is unclear in a free fermion since a unique boundary condition is allowed in $\mathbb{H}^d$ \cite{Dowker:1995sw,Rodriguez-Gomez:2017aca}.

The organisation of this paper is as follows.
In the next section, we summarise coordinate systems and Weyl transformations among the coordinate systems.
Furthermore, we classify boundary conditions of a massive fermion on $\mathbb{H}^d$ and 
a massless fermion on $\mathbb{H}^{p+1} \times \mathbb{S}^{q-1}$. 
In particular, for a massless fermion on $\mathbb{H}^{p+1} \times \mathbb{S}^{q-1}$,
we show that a boundary condition of a Dirichlet type always exists while a boundary condition of a Neumann type exists only in $q=2$.
In section \ref{sec3}, we compute free energies of a massless fermion on $\mathbb{S}^d$ and a hemisphere $\mathbb{HS}^d$ using a zeta-function regularisation for a warm-up of the next section.
In section \ref{sec4}, we compute free energies of a massive fermion on $\mathbb{H}^d$ and free energies of a massless fermion $\mathbb{H}^{p+1} \times \mathbb{S}^{q-1}$ with Dirichlet boundary condition by using a zeta-function regularisation.
In section \ref{sec5}, we obtain free energies with Neumann boundary condition by analytical continuation and confirm the validity of our conjecture.
The final section is devoted to discussion.
Appendix \ref{app_table} contains the list of anomaly parts and finite parts of free energies on $\mathbb{S}^d$, $\mathbb{HS}^d$, $\mathbb{H}^d$ and $\mathbb{H}^{p+1} \times \mathbb{S}^{q-1}$ with Dirichlet boundary condition, and appendix \ref{app2} is a technical details of a computation. 

\section{Classification of boundary conditions}
\label{sec2}

\subsection{Coordinate system}

In this section, we summarise coordinate systems for a sphere $\mathbb{S}^d$, a hemisphere $\mathbb{HS}^d$, a hyperbolic space $\mathbb{H}^d$ and $\mathbb{H}^{p+1} \times \mathbb{S}^{q-1}$ and conformal maps among a flat space $\mathbb{R}^d$ and them.

Let us consider DCFT$_d$ on  $\mathbb{R}^d$,
\begin{align}
    \dd s^2 = \dd x_a^2 + \dd y_i^2 \,, \qquad (a=1,\cdots, p, i = p+1,\cdots,d)
\end{align}
where a $p$-dimensional defect sits at $y_i = 0$.
For later convenience, we introduce a codimension of the defect,
\begin{align}
    q= d-p \,.
\end{align}
By using the polar coordinate for the $y_i$-coordinate, 
\begin{align}
    \dd y_i^2 = \dd z^2 + z^2 \dd s_{\mathbb{S}^{q-1}}^2 \,, 
\end{align}
the metric of the flat space becomes
\begin{align}\label{Flat_defect_map}
    \begin{aligned}
    \dd s^2 & = \dd x_a^2 + \dd z^2 + z^2 \dd s_{\mathbb{S}^{q-1}}^2 \\
    & = z^2 \left( \frac{\dd x_a^2 + \dd z^2}{z^2} + \dd s_{\mathbb{S}^{q-1}}^2 \right) \,.
    \end{aligned}
\end{align}
After a Weyl rescaling, the metric \eqref{Flat_defect_map} reduces to a geometry $\mathbb{H}^{p+1} \times \mathbb{S}^{q-1}$ with radius $R$,
\begin{align} \label{metric_poincare}
    \dd s^2 = R^2 \left( \frac{\dd x_a^2 + \dd z^2}{z^2} +  \dd s_{\mathbb{S}^{q-1}}^2 \right) \,.
\end{align}
Now the defect sits at the boundary of $\mathbb{H}^{p+1}$.
We can also use the global coordinate for $\mathbb{H}^{p+1}$,
\begin{align} \label{metric_hyperbolic}
  \dd s^2 = R^2 \left( \dd \rho^2 +\sinh^2 \rho \, \dd s_{\mathbb{S}^p}^2 + \, \dd s_{\mathbb{S}^{q-1}}^2 \right) \,,
\end{align}
where the defect sits at $\rho=\infty$.
Introducing a new variable $\varphi$, $\tan \varphi = \sinh \rho$, the metric \eqref{metric_hyperbolic} becomes
\begin{align} \label{metric2.6}
    \dd s^2 = \frac{R^2}{\cos^2 \varphi} \left( \dd \varphi ^2 +\sin^2 \varphi \, \dd s_{\mathbb{S}^p}^2+ \cos ^2 \varphi \, \dd s_{\mathbb{S}^{q-1}}^2 \right) \,.
\end{align}
After a Weyl rescaling, the metric \eqref{metric2.6} can be mapped to the sphere metric with radius $R$
\begin{align} \label{metric_sphere}
    \dd s^2 = R^2 \left( \dd \varphi ^2 +\sin^2 \varphi \, \dd s_{\mathbb{S}^p}^2+ \cos ^2 \varphi \, \dd s_{\mathbb{S}^{q-1}}^2 \right) \,,
\end{align}
where $0 \leq \varphi < \pi$ and the defect sits at $\varphi = \pi/2$. 
The hemisphere $\mathbb{HS}^d$ has the same metric of the sphere \eqref{metric_sphere}.
A different point is that the range of $\varphi$ is $0 \leq \varphi \leq \pi/2$ and the boundary sits at $\varphi = \pi/2$.

\subsection{Boundary condition of fermion on \texorpdfstring{$\mathbb{H}^{p+1} \times \mathbb{S}^{q-1}$}{HxS}}
\label{sec2.2}

In this section, we classify allowed boundary conditions of a massless fermion on $\mathbb{H}^{p+1} \times \mathbb{S}^{q-1}$.
A free fermion on a curved background which is conformally equivalent to $\mathbb{R}^d$ is studied in e.g. \cite{Camporesi:1992tm,Camporesi:1995fb,Lewkowycz:2012qr,Herzog:2019bom,Klebanov:2011uf}.
See \cite{Chester:2015wao} for the notation of a fermion on $\mathbb{H}^2 \times \mathbb{S}^2$.
The action of a massive Dirac fermion is given by\footnote{The massive fermion is not conformal. 
However, we introduce a mass term for later convenience.}
\begin{align}
    I = \int \! \dd^d x \sqrt{g} \, \left( \i\,  \psi^\dagger \, \Gamma^a \nabla_a \psi +M\psi^\dagger \psi \right) \,,
\end{align}
where we assume $M \geq 0$.
The rank of the gamma matrix for $d\geq 2$ is 
\begin{align} \label{rank}
    r_d = 2^{\floor{\frac{d}{2}}} \,,
\end{align}
and the gamma matrix satisfies the anti-commutation relation,
\begin{align}
    \{ \Gamma^a , \Gamma^b \} = 2\delta^{ab} \mathds{1} \,.
\end{align}
The covariant derivative $\nabla_a$ and the spin connection $\omega_{\mu bc}$ are defined as 
\begin{align}
\begin{aligned}
    \nabla_a & = e_a^\mu \nabla_\mu \,, \qquad 
    \nabla_\mu = \partial_\mu + \frac{1}{2} \sigma^{bc} \omega_{\mu bc} \,, \\
    \sigma^{ab} &= \frac{1}{4} [\Gamma^a,\Gamma^b]\,, \qquad 
    \omega_{\mu bc} = e_b^\nu (\partial_\mu e_{c\nu} -\Gamma_{\nu \mu}^\alpha e_{c\alpha}) 
\end{aligned}
\end{align}
using a frame field $e_a^\mu$, which satisfies
\begin{align}
    e_a^\mu e_b^\nu g_{\mu \nu}= \delta_{ab} \,.
\end{align}
The covariant derivative satisfies a relation,
\begin{align}
    ( \Gamma^a \nabla_a )^2 = \nabla^2 -\frac{1}{4}\mathcal{R} \,,
\end{align}
where $\mathcal{R}$ is a Ricci scalar.

For $\mathbb{H}^d$, a solution of the equation of motion for the massive fermion behaves 
\begin{align}
    \psi \sim z^{\Delta_\pm}
\end{align}
near the boundary at $z=0$,
where $\Delta_\pm$ is given by 
\begin{align}
     \Delta_\pm - \frac{d-1}{2} = \pm MR \,.
\end{align}
Then, $\Delta_\pm$ become degenerate in the massless limit, and the allowed asymptotic behaviour is unique for a massless fermion.

For the product space $\mathbb{H}^{p+1}\times \mathbb{S}^{q-1}$, we consider a massless fermion from the beginning.
We first decompose the fermionic field as 
\begin{align} \label{field_decomposition}
    \psi (z,x,\theta) = \sum_\ell \psi_{\mathbb{H}^{p+1}} (z,x) \otimes \psi_{\ell,\mathbb{S}^{q-1}} (\theta) \,, 
\end{align}
and the gamma matrices are also decomposed similarly.
For even $p$ and even $q \geq 4$, the decomposition of the fermion \eqref{field_decomposition} has an additional $U(1)$ as is clear from that the rank of the spinor representation are different in both sides of \eqref{field_decomposition}.\footnote{We are indebted to D.~Rodriguez-Gomez and J.~G.~Russo for the decomposition of the fermion.}
Mathematically, this is equivalent to a decomposition of a representation of $SO(\frac{p+q}{2})$ to that of $SO(\frac{p}{2})\times SO(\frac{q-2}{2}) \times U(1)$.
The spherical part $\psi_{\ell,\mathbb{S}^{q-1}} (\theta)$ satisfies the equation,
\begin{align}
    (\Gamma \cdot \nabla)_{\mathbb{S}^{q-1}} \psi_{\ell,\mathbb{S}^{q-1}} (\theta) = \pm \, \i \, \frac{\ell + \frac{q-1}{2}}{R} \psi_{\ell,\mathbb{S}^{q-1}} (\theta) \,,
\end{align}
where we decomposed the covariant derivative on $\mathbb{H}^{p+1} \times \mathbb{S}^{q-1}$ into the covariant derivatives on $\mathbb{H}^{p+1}$ and $\mathbb{S}^{q-1}$ appropriately.
Then, the equation of motion reduces to 
\begin{align}
    \left(\i \, (\Gamma \cdot \nabla)_{\mathbb{H}^{p+1}} \pm \frac{\ell +\frac{q-1}{2}}{R} \right) \psi_{\mathbb{H}^{p+1}} (z,x) =0 \,.
\end{align}
The solution of this equation of motion behaves as 
\begin{align}
    \psi_{\mathbb{H}^{p+1}} (z,x) \sim z^{\Delta_\pm^\ell}
\end{align}
near the boundary, $z=0$, and $\Delta_\pm^\ell$ is given by 
\begin{align}\label{defect_op_dimension}
    \Delta_\pm^\ell = \frac{p}{2} \pm \left( \ell + \frac{q-1}{2} \right) \,.
\end{align}
The parameter $\Delta_\pm^\ell$ can be understood as the conformal dimensions of operators localising on a $p$-dimensional conformal defect at the boundary of $\BH^{p+1}$.
Then not all the operators with conformal dimensions \eqref{defect_op_dimension} are allowed to exist due to the unitarity bound in $p$ dimensions \cite{Minwalla:1997ka}:
\begin{align} \label{unitarity_bound}
    \Delta \geq \frac{p-1}{2} \,.
\end{align}
$\Delta_+^\ell$ is always above the unitarity bound, while $\Delta_-^\ell$ is not necessary to satisfy the unitarity bound unless
\begin{align}\label{bound_ell}
    \ell \leq 1 - \frac{q}{2} \ .
\end{align}
Hence the mode with $\ell=0$ for $q=2$ is allowed to have the boundary conditions corresponding to $\Delta_-^\ell$.

The allowed boundary conditions for the massless fermion are classified as follows and are listed in table \ref{tab:classification}.

\paragraph{$q=1$ case:}
In this case, the geometry is the hyperbolic space $\mathbb{H}^d$, and the allowed asymptotic behaviour is unique,
\begin{align}
    \text{Mixed b.c.}: \quad \Delta = \frac{d-1}{2} \,.
\end{align}
This boundary condition is called a mixed boundary condition.

\paragraph{$q=2$ case:}
    Only the $\ell=0$ mode is allowed, resulting in a nontrivial boundary condition with $\Delta^{\ell=0}_-= \frac{p-1}{2}$ for $p\ge 2$. Note that $\Delta^{\ell=0}_-$ saturates the unitarity bound.
    Then, two different boundary conditions are allowed, 
    \begin{align}
    \begin{aligned}
        \text{Dirichlet b.c.}:& \quad  \Delta_\text{D} = \Delta_+^\ell \quad \text{for all } \ell \,, \\
        \text{Neumann b.c.}:& \quad  \Delta_\text{N} = 
        \begin{dcases}
           \Delta_-^{\ell} & \quad \text{for } \ell = 0\,, \\
           \Delta_+^\ell & \quad \text{for } \ell \neq 0\,.
        \end{dcases}
    \end{aligned}
    \end{align}
In particular,  $\Delta^{\ell=0}_-$ vanishes for $p=1$ case, and this implies that the defect operator is the identity operator. 
Thus, we exclude $p=1$ case of a nontrivial boundary condition.

\paragraph{$q \geq 3$ case:}
Only the Dirichlet type boundary condition is allowed,
    \begin{align}
        \text{Dirichlet b.c.}: \quad  \Delta_\text{D} = \Delta_+^\ell \quad \text{for all } \ell \,.
    \end{align}

\begin{table}[ht]
    \centering
    \begin{tabularx}{\linewidth}{cYYYYYY}
        \toprule
          & $q=1$ & $q=2$ & $ q=3$ & $q=4$ & $q=5$ &  $\cdots$ \\ \hline
         $p=1$ &  $\Delta$ & $\Delta_\text{D}$ & $\Delta_\text{D}$ & $\Delta_\text{D}$ & $\Delta_\text{D}$ & $\cdots$ \\
         $p=2$ &  $\Delta$ & \cellcolor{gray!10} $\Delta_\text{D}/\Delta_-^{\ell=0}$ &  $\Delta_\text{D}$ & $\Delta_\text{D}$ & $\Delta_\text{D}$ & \\
         $p=3$ & $\Delta$ & \cellcolor{gray!10} $\Delta_\text{D}/\Delta_-^{\ell=0}$ &  $\Delta_\text{D}$ & $\Delta_\text{D}$ & $\Delta_\text{D}$ &  $\cdots$ \\
         $p=4$ &  $\Delta$ & \cellcolor{gray!10} $\Delta_\text{D}/\Delta_-^{\ell=0}$ &  $\Delta_\text{D}$ &  $\Delta_\text{D}$ & $\Delta_\text{D}$ &  \\
         $p=5$ & $\Delta$ & \cellcolor{gray!10} $\Delta_\text{D}/\Delta_-^{\ell=0}$ & $\Delta_\text{D}$ & $\Delta_\text{D}$ & $\Delta_\text{D}$ &  \\
         $\vdots$ & $\vdots$ & \cellcolor{gray!10} & & $\vdots$ & &  $\ddots$ \\
         \bottomrule
    \end{tabularx}
    \caption{Classification of the allowed boundary conditions in the free massless fermion.
    The Neumann boundary conditions 
    exist in the shaded cells and the allowed modes differ from the Dirichlet ones are shown in the right side. For $q=1$, the boundary condition is unique. For $q=2$, $\Delta_-^{\ell =0}$ saturates the unitarity bound \eqref{unitarity_bound}.
    }
    \label{tab:classification}
\end{table}

\section{Free energy on \texorpdfstring{$\mathbb{S}^d$}{Sd} and \texorpdfstring{$\mathbb{HS}^d$}{HSd}}
\label{sec3}

In this section, we compute free energies of a free massless Dirac fermion on $\mathbb{S}^d$ and $\mathbb{HS}^d$ using a zeta-function regularisation for a warm-up of the next section.

\subsection{Free energy on \texorpdfstring{$\BS^d$}{Sd}}

For a massless fermion on $\BS^d$, 
the free energy is given by\footnote{As noted in \cite{Nishioka:2021uef,Monin:2016bwf}, there is an ambiguity to decompose the logarithmic function into two parts. We require two conditions to remove the ambiguity:  the free energy in the Schwinger representation should be convergent and the resulting zeta function does not depend on the cutoff scale $\tilde \Lambda$.
In \eqref{unrenom_F_al}, the two  decomposed  logarithmic functions are the same and the free energy in the Schwinger representation \eqref{sphere_FE_schwinger} is convergent in the $\ell \to \infty$ limit.}
\begin{align} \label{unrenom_F_al}
    \begin{aligned}
        F[\BS^d] 
            &= - \frac{1}{2} \text{tr} \log \left[ -\tilde{\Lambda}^{-2}  (\Gamma^a \nabla_a)^2  \right] \\
            &= - \sum_{\ell =0}^\infty g^{(d)}(\ell)\,  \log \left( \frac{\nu_\ell^{(d)}}{\tilde\Lambda R}\right) \,,
    \end{aligned}
\end{align}
where we use the equation, 
\begin{align}
    \Gamma^a \nabla_a \psi_\ell = \pm \, \i \, \frac{\nu_\ell^{(d)}}{R} \psi_\ell 
\end{align}
with the eigenvalues,
\begin{align} 
        \nu_\ell^{(d)} = \ell + \frac{d}{2} \,, \qquad \ell = 0,1,2, \cdots \,.
\end{align}
The degeneracy for each sign is given by 
\begin{align} \label{degeneracy_half}
        g_\pm^{(d)}(\ell)  = \frac{r_d \Gamma (\ell +d)}{\Gamma (d) \Gamma (\ell +1 ) } \,, 
\end{align}
where $r_d$ \eqref{rank} is the rank of the gamma matrix as before.
For later convenience, we introduce a notation 
\begin{align} \label{degeneracy}
        g^{(d)}(\ell)  = 2 g_\pm^{(d)}(\ell) =  \frac{2 r_d \Gamma (\ell +d)}{\Gamma (d) \Gamma (\ell +1 ) } \,.
\end{align}

We write the free energy \eqref{unrenom_F_al} in the Schwinger representation,
\begin{align} \label{sphere_FE_schwinger}
    F[\BS^d] =  \int_0^\infty\,\frac{\dd t}{t}\,\sum_{\ell =0}^\infty g^{(d)}(\ell)\, \mathrm{e}^{-t \nu_\ell^{(d)}/(\tilde\Lambda R)} \,,
\end{align}
and we introduce the regularised free energy \cite{Vassilevich:2003xt} to remove the divergence in the integral,
\begin{align}
    \begin{aligned}
        F_{s}[\BS^d] &=              \int_0^\infty\,\frac{\dd t}{t^{1-s}}\,\sum_{\ell =0}^\infty g^{(d)}(\ell) \, \mathrm{e}^{-t \nu_\ell^{(d)}/(\tilde\Lambda R)} \\
            &=  \frac{1}{2} (\tilde\Lambda R)^{s}\,\Gamma(s)\, \zeta_{\BS^d} (s) \,,
    \end{aligned}
\end{align}
where the zeta function $\zeta_{\BS^d} (s)$ is defined by
\begin{align}
    \zeta_{\BS^d}(s) \equiv 2 \sum_{\ell =0}^\infty g^{(d)}(\ell)\, \left( \nu_\ell^{(d)} \right)^{-s} \ .
\end{align}
Then the (unregularised) free energy is obtained in the $s\to 0$ limit:
\begin{align}
    F_s [\BS^d] =  \frac{1}{2} \left( \frac{1}{s} - \gamma_\text{E} + \log(\tilde\Lambda R)\right)\, \zeta_{\BS^d}(0) + \frac{1}{2} \partial_s \zeta_{\BS^d}(0) + \mathcal{O}(s) \ ,
\end{align}
which is divergent due to the pole at $s=0$.
Here $\gamma_\text{E}$ is the Euler constant.
By removing the pole, the remaining part becomes the renormalized free energy
\begin{align}\label{Fren_sphere}
    F_\text{ren}[\BS^d] \equiv  \frac{1}{2} \partial_s \zeta_{\BS^d}(0) + \frac{1}{2} \log(\Lambda R)\, \zeta_{\BS^d}(0) \ ,
\end{align}
where $\Lambda = \mathrm{e}^{-\gamma_\text{E}}\,\tilde\Lambda$.

To compute the zeta function, we expand the gamma functions in the degeneracy \eqref{degeneracy}, 
\begin{align} \label{expansion}
    \frac{\Gamma \left(\nu_\ell^{(d)} +\frac{d}{2} \right) }{ \Gamma \left(\nu_\ell^{(d)} - \frac{d}{2}+1 \right)} =
\begin{dcases}
        \sum_{n=0}^{\frac{d}{2}}(-1)^{\frac{d}{2}+n}\,\alpha_{n, d+1}\,\left(\nu_\ell^{(d)}\right)^{2n-1} & d: \text{even}\,, \\
         \sum_{n=0}^{\frac{d-1}{2}}(-1)^{\frac{d-1}{2}+n}\,\beta_{n, d+1}\,\left(\nu_\ell^{(d)}\right)^{2n} & d: \text{odd} \,,
    \end{dcases}
\end{align}
where we used the awkward suffix for $\alpha_{n, d+1}$ and $\beta_{n, d+1}$ to use the same notation in the scalar case \cite{Nishioka:2021uef}.
Since $\alpha_{0,d+1} = 0$, we can omit the $n=0$ term in the summation for even $d$ whenever we want.
Using the asymptotic expansion ((5.11.14) in \cite{NIST:DLMF})
\begin{align}
    \frac{\Gamma(x + a)}{\Gamma(x + b)} = \sum_{k=0}^\infty\,\left(x + \frac{a + b - 1}{2}\right)^{a-b-2k}\,\binom{a-b}{2k}\,B_{2k}^{(a-b+1)}\left(\frac{a-b+1}{2}\right) \ ,
\end{align}
where $B_k^{(m)}(x)$ is the generalized Bernoulli polynomial, and comparing both sides, we find
\begin{align}\label{coeff_alpha}
    \text{Even } d: \qquad \alpha_{n,d+1} &= (-1)^{\frac{d}{2}+n}\, \binom{d-1}{d-2n}\, B^{(d)}_{d-2n} \left( \frac{d}{2}\right) \,, \\
\label{coeff_beta}
    \text{Odd } d: \qquad \beta_{n, d+1} &= (-1)^{\frac{d-1}{2} + n}\, \binom{d-1}{d-1 - 2n}\,B^{(d)}_{d-1-2n} \left( \frac{d}{2}\right) \,.
\end{align}

\subsubsection{Odd \texorpdfstring{$d$}{d}}
When $d$ is odd, the zeta function reduces to
\begin{align}
    \zeta_{\mathbb{S}^d} (s) 
    = \frac{4r_d}{\Gamma (d)}   \sum_{n=0}^{\frac{d-1}{2}} (-1)^{\frac{d-1}{2}+n} \beta_{n,d+1}\, \zeta_\text{H} \left( s-2n, \frac{1}{2} \right) \,,
\end{align}
where we use the identity
\begin{align}
    \zeta_\text{H} \left(s-2n,\frac{d}{2} \right)   = \zeta_\text{H} \left(s-2n,\frac{1}{2} \right) -     \sum_{m=0}^{\frac{d-1}{2}}
    \left( m+\frac{1}{2}\right)^{2n-s}
\end{align}
and the expansion \eqref{expansion} with the replacement $\nu_\ell^{(d)} \to m+1/2$.
Since the zeta function at $s=0$ vanishes,
\begin{align}
\begin{aligned}
    \zeta_{\mathbb{S}^d} (0) 
        &=
        \frac{4r_d}{\Gamma (d)}   \sum_{n=0}^{\frac{d-1}{2}} (-1)^{\frac{d-1}{2}+n} \beta_{n,d+1}\, \zeta_\text{H} \left( -2n, \frac{1}{2} \right) \\
        &= 0 \,,
    \end{aligned}
\end{align}
due to the fact $\zeta_\text{H} \left( -2n, 1/2 \right)=0$, there is no conformal anomaly, and the finite part remains in the free energy:
\begin{align} \label{finite_sphere}
    \begin{aligned}
    F_\text{ren} [\BS^d]
        &= \frac{1}{2} \partial_s \zeta_{\BS^d}(0) \\
        & =  \frac{r_d}{\Gamma (d)}  (-1)^{\frac{d+1}{2}} \beta_{0,d+1}\, \log 2
        +\frac{2r_d}{\Gamma (d)}   \sum_{n=1}^{\frac{d-1}{2}} (-1)^{\frac{d-1}{2}+n} \beta_{n,d+1}\, (2^{-2n}-1) \zeta' (-2n) \,.
    \end{aligned}
\end{align}
This formula correctly reproduces the known results \cite{Marino:2011nm,Klebanov:2011gs}.
The finite parts of the free energy for $d \leq 9$ are listed in table \ref{tab:Fin_p_q} in appendix \ref{app_table}.

\subsubsection{Even \texorpdfstring{$d$}{d}}

When $d$ is even the zeta function is given by 
\begin{align}
    \zeta_{\BS^d}(s) 
    = \frac{4r_d}{\Gamma (d)} \sum_{n=0}^{\frac{d}{2}}(-1)^{\frac{d}{2}+n}\,\alpha_{n, d+1}\, \zeta ( s-2n+1 ) \,,
\end{align}
where we use the identity
\begin{align}
    \begin{aligned}
    \zeta_\text{H} \left(s-2n+1,\frac{d}{2} \right)   = \zeta (s-2n+1)  -     \sum_{m=0}^{\frac{d}{2}-2}
    \left( m+1 \right)^{2n-1-s}
    \end{aligned}
\end{align}
and the expansion \eqref{expansion} with the replacement $\nu_\ell^{(d)} \to m+1$.
The renormalized free energy is given by
\begin{align} \label{sphere_FE_even}
    F_\text{ren}[\BS^d] 
        = 
         - A[\BS^d] \,\log(\Lambda R) + F_{\text{fin}}[\BS^d]  \ ,
\end{align}
with the anomaly coefficient
\begin{align} \label{anomaly_sphere}
\begin{aligned}
        A[\BS^d] & = - \frac{1}{2} \zeta_{\BS^d}(0) \\ 
   & = \frac{r_d}{\Gamma (d)} \sum_{n=1}^{\frac{d}{2}}(-1)^{\frac{d}{2}+n}\,\alpha_{n, d+1}\, \frac{B_{2n}}{n} \,,
\end{aligned}
\end{align}
where $B_{2n}$ is the Bernoulli number, and the finite part
\begin{align}
\begin{aligned}
    F_{\text{fin}}[\BS^d]& =  \frac{1}{2} \partial_s  \zeta_{\BS^d}(0) \\
    & = \frac{2r_d}{\Gamma (d)} \sum_{n=1}^{\frac{d}{2}}(-1)^{\frac{d}{2}+n}\,\alpha_{n, d+1}\, \zeta' ( -2n+1 ) \,.
\end{aligned}
\end{align}
There is a logarithmic divergent term associated with the conformal anomaly in the free energy.
The free energy \eqref{sphere_FE_even} with \eqref{anomaly_sphere} correctly reproduces the known conformal anomaly \cite{Giombi:2014xxa}. 
The anomaly coefficients and the finite parts of the free energy for $d \leq 10$ are listed in tables \ref{tab:F_p_q} and \ref{tab:Fin_p_q} in appendix \ref{app_table}.

\subsubsection{Interpolating \texorpdfstring{$a$}{a} and \texorpdfstring{$F$}{F}}

The finite part of the free energy \eqref{finite_sphere} and the anomaly coefficient of the free energy \eqref{anomaly_sphere} do not depend on the choice of the cutoff scale and are universal in this sense.
Thus, we introduce the ``universal" free energy:
\begin{align} \label{free_energy_univ}
    F_{\text{univ}}[\BS^{d}]  =
    \begin{dcases}
            F_\text{fin}[\BS^{d}] &\qquad d: \text{odd} \,, \\
            -A [\BS^{d}]\, \log \left( \frac{R}{\epsilon}\right)  &\qquad d: \text{even} \,,
    \end{dcases}
\end{align}
where $\epsilon$ is used for the cutoff instead of $\Lambda$. 
In \cite{Giombi:2014xxa}, it is pointed out that the universal free energy has an integral representation:
\begin{align} \label{integral_expression}
    F_\text{univ} [\mathbb{S}^d] 
    = - \frac{2 r_d }{\sin \left( \frac{\pi d}{2} \right)\Gamma (d+1)} \int_0^\frac{1}{2} \! \dd u \, \cos\left( \pi u \right) \Gamma \left( \frac{d+1}{2}+u \right) \Gamma \left( \frac{d+1}{2}-u \right) \,.
\end{align}
For odd $d$, the prefactor of the universal free energy \eqref{integral_expression} is finite.
However, for even $d$, the prefactor in \eqref{integral_expression} is divergent due to the sine function.
This divergence may be replaced with the logarithmic divergence by introducing a small cutoff $\epsilon$,
\begin{align} \label{replacement2}
    -\frac{1}{\sin\left(\frac{\pi d}{2}\right)} =
    \begin{dcases}
            (-1)^\frac{d+1}{2}  &\qquad d: \text{odd} \,, \\
            (-1)^\frac{d}{2}\, \frac{2}{\pi} \,\log \left( \frac{R}{\epsilon}\right) &\qquad d: \text{even} \,.
    \end{dcases}
\end{align}
Using the replacement \eqref{replacement2}, the universal free energy in the integral representation \eqref{integral_expression} has the same behaviour of \eqref{free_energy_univ}.
A proof of the equivalence of the two expressions \eqref{free_energy_univ}
and \eqref{integral_expression} is presented in \cite{Dowker:2017cqe}.

\subsection{Free energy on \texorpdfstring{$\mathbb{HS}^d$}{HSd}}

Next, let us consider the free energy on the hemisphere.
At the boundary of $\mathbb{HS}^d$, a mixed boundary condition is imposed \cite{Dowker:1995sw},
\begin{align}
    P_+ \psi = 0 \,.
\end{align} 
Here $P_+$ is a projection operator,
\begin{align}
    P_+ = \frac{1}{2} \left( 1 - \i \,  \Gamma^\ast \Gamma^a e_a^\mu n_\mu  \right)
\end{align}
with a chirality matrix $\Gamma^\ast$ and an incoming normal vector $n_\mu$.
See appendix A in \cite{Dowker:1995sw} for the detail of a construction of the chirality matrix $\Gamma^\ast$.
The mixed boundary condition preserves a conformal symmetry.

A degeneracy with the mixed boundary condition is given by \eqref{degeneracy_half} which is just a half of the degeneracy of $\mathbb{S}^{d}$ \eqref{degeneracy}.
Then, the free energy on $\mathbb{HS}^d$ is nothing but half of that on $\mathbb{S}^d$,
\begin{align}
    F_\text{ren} [\mathbb{HS}^d] = \frac{1}{2} F_\text{ren} [\mathbb{S}^d] \,.
\end{align}
This formula correctly reproduces the known results \cite{Rodriguez-Gomez:2017aca}.\footnote{In \cite{Rodriguez-Gomez:2017aca}, anomaly coefficients of a ball $\mathbb{B}^d$ which is conformally equivalent to $\mathbb{H}^d$, is obtained.}

\section{Free energy on \texorpdfstring{$\mathbb{H}^{p+1} \times \mathbb{S}^{q-1}$}{H x S} with Dirichlet boundary condition}
\label{sec4}

In this section, we first compute a free energy of a massive fermion on $\mathbb{H}^d$, although we are interested in a massless (conformal) fermion.
This is because the angular momentum along $\mathbb{S}^{q-1}$ can be regarded as a mass on $\mathbb{H}^{p+1}$ due to a Kaluza-Klein mechanism when we compute the free energy on $\mathbb{H}^{p+1} \times \mathbb{S}^{q-1}$.
After that, we compute a free energy of a massless fermion on $\mathbb{H}^{p+1} \times \mathbb{S}^{q-1}$.

\subsection{Free energy on \texorpdfstring{$\mathbb{H}^d$}{Hd}}
\label{sec4.1}

In this section, we extend a computation of the zeta function for a massive fermion on $\BH^d$ for $d=3,4$ in \cite{Bytsenko:1994bc} to general dimensions.

The free energy of the massive Dirac fermion with mass $M=m/R$ is given by 
\begin{align} \label{free_energy_hyperbolic}
\begin{aligned}
        F[\mathbb{H}^d] (m) &= - \frac{1}{2} \text{tr} \log \left[ -\tilde{\Lambda}^{-2} \left( (\Gamma^a \nabla_a)^2 - M^2 \right) \right] \\
    & =- \frac{1}{2}  \int_0^\infty \! \dd \omega \, \mu^{(d)} (\omega) \left[ \log \left(\frac{\omega+\i \,  m}{\tilde{\Lambda}R} \right) +\log \left(\frac{\omega -\i \, m}{\tilde{\Lambda}R} \right) \right] \,,
\end{aligned}
\end{align}
where $\tilde\Lambda$ is the UV cutoff scale introduced to make the integral dimensionless.
The parameter $\omega$ is an eigenvalue of the equation, 
\begin{align}
    \Gamma^a \nabla_a \psi_\omega = \i \, \frac{\omega}{R} \psi_\omega \,, \qquad \omega \geq 0 \,,
\end{align}
and the Plancherel measure of a fermion on $\mathbb{H}^d$ of unit radius takes the form \cite{Camporesi:1995fb}
\begin{align} \label{Plancherel}
\begin{aligned}
    \mu^{(d)} (\omega) &
    = c_d \, r_d \, \left| \frac{\Gamma \left(\frac{d}{2}+ \i \, \omega \right)}{\Gamma \left(\frac{1}{2}+\i \, \omega \right)} \right|^2 \\
    & = c_d \, r_d \, 
    \begin{dcases}
        \prod_{j=\frac{1}{2}}^\frac{d-2}{2}(\omega^2 + j^2) & d: \text{odd} \,, \\
            \omega\,\coth (\pi\omega)\,\prod_{j=1}^\frac{d-2}{2}(\omega^2 + j^2) & d: \text{even} \,,
            \end{dcases}
\end{aligned}
\end{align}
with the coefficient 
\begin{align}
    c_d = \frac{\text{Vol}(\mathbb{H}^d) }{2^{d-1} \pi^{d/2}\Gamma (d/2)}
\end{align}
and the rank of the gamma matrix $r_d$ \eqref{rank}.
The regularised volume of the hyperbolic space is given by 
\begin{align}
    \text{Vol} (\mathbb{H}^d) = \pi^{\frac{d-1}{2}} \Gamma \left( \frac{1-d}{2} \right) = - \frac{\pi^{\frac{d+1}{2}} }{\sin \left( \pi \frac{d-1}{2} \right) \Gamma \left( \frac{d+1}{2} \right)} \,.
\end{align}
The hyperbolic volume is finite for even $d$ but divergent for odd $d$ due to the pole of the sine function.
By introducing a small cutoff parameter $\epsilon$, the sine function can be replaced by the logarithmic divergence,
\begin{align} \label{replacement}
    -\frac{1}{\sin\left(\pi\,\frac{d-1}{2}\right)} =
    \begin{dcases}
            (-1)^\frac{d-1}{2} \frac{2}{\pi} \,\log \left( \frac{R}{\epsilon}\right) &\qquad d: \text{odd} \,, \\
            (-1)^\frac{d}{2}  &\qquad d: \text{even} \,.
    \end{dcases} 
\end{align}
Then, the coefficient becomes
\begin{align}
    c_d = - \frac{1}{\sin \left( \pi \frac{d-1}{2} \right) \Gamma (d)} = \frac{1}{\Gamma (d)} \begin{dcases}
            (-1)^\frac{d-1}{2}\, \frac{2}{\pi} \,\log \left( \frac{R}{\epsilon}\right) &\qquad d: \text{odd} \,, \\
            (-1)^\frac{d}{2} &\qquad d: \text{even} \,.
    \end{dcases}
\end{align}
Since the Plancherel measure \eqref{Plancherel} needs to satisfy the square integrability condition, the free energy is defined for $m \geq 0$. That is, equation \eqref{free_energy_hyperbolic} represents a free energy with a boundary condition, $\Delta = (d-1)/2+m$.
We add a mass term in the free energy \eqref{free_energy_hyperbolic} for the regularisation of the zero mode, and we take a massless limit to obtain free energies of the conformal fermion.

By using the Schwinger representation, the free energy can be written as  
\begin{align}
    F[\mathbb{H}^d] (m) = \frac{1}{2} \int_{0}^\infty \! \frac{\dd t}{t} \int_0^\infty \! \dd \omega \, \mu^{(d)} (\omega) \, \left( \mathrm{e}^{ -t (\omega +\i \,  m)/\tilde{\Lambda}R} + \mathrm{e}^{ -t (\omega -\i \, m)/\tilde{\Lambda}R} \right) \,.
\end{align}
To remove the divergence in the integral, we introduce the regularised free energy
\begin{align}
\begin{aligned}
         F_s[\mathbb{H}^d] (m) & = \frac{1}{2} \int_{0}^\infty \! \frac{\dd t}{t^{1-s}} \int_0^\infty \! \dd \omega \, \mu^{(d)} (\omega) \, \left(\mathrm{e}^{ -t (\omega +\i \,  m)/\tilde{\Lambda}R} + \mathrm{e}^{ -t (\omega -\i \, m)/\tilde{\Lambda}R} \right) \\
    & = \frac{1}{2} (\tilde{\Lambda}R)^{s} \Gamma (s) \zeta_{\mathbb{H}^d} (s,m) \,,
\end{aligned}
\end{align}
where the zeta function is defined as 
\begin{align}
    \zeta_{\mathbb{H}^d} (s,m) = \int_0^\infty \! \dd \omega \, \mu^{(d)} (\omega) \, \left( (\omega+\i \, m)^{-s} + (\omega - \i \, m)^{-s} \right) \,.
\end{align}
Then, the (unregularised) free energy is obtained in the $s\to 0$ limit,
\begin{align}
    F_s [\mathbb{H}^d] (m) =  \frac{1}{2} \left( \frac{1}{s}-\gamma_\text{E} +\log (\tilde{\Lambda}R) \right) \zeta_{\mathbb{H}^d} (s,m) +\frac{1}{2} \partial_s \zeta_{\mathbb{H}^d} (s,m) + \mathcal{O} (s) \,.
\end{align}
By removing the pole at $s=0$, we obtain the renormalized free energy
\begin{align}
    F_{\text{ren}}[\mathbb{H}^d] (m) =  \frac{1}{2} \partial_s \zeta_{\mathbb{H}^d} (0,m) + \frac{1}{2} \log (\Lambda R)  \zeta_{\mathbb{H}^d} (0,m) \,,
\end{align}
where $\Lambda = \mathrm{e}^{-\gamma_\text{E}}\tilde{\Lambda}$.

In the following, we will compute the renormalized free energy by evaluating the zeta function based on the method used in \cite{Camporesi:1994ga,Bytsenko:1995ak}.

\subsubsection{Odd \texorpdfstring{$d$}{d}}
\label{sec4.1.1}

Using the expansion of the Plancherel measure \eqref{Plancherel}
\begin{align}
    \mu^{(d)} (\omega) 
    = c_d \, r_d \sum_{k=0}^{\frac{d-1}{2}} \beta_{k,d+1} \omega^{2k} \,,
\end{align}
the zeta function is convergent for $\text{Re}\, s > 2k+1$,
\begin{align}
\begin{aligned}
    \zeta_{\mathbb{H}^d} (s,m) & = c_d \, r_d \sum_{k=0}^{\frac{d-1}{2}} \beta_{k,d+1} \int_0^\infty \! \dd \omega \,  \omega^{2k} \left((\omega + \i \, m)^{-s} + (\omega - \i \,  m)^{-s} \right) \\
    & = 2c_d \, r_d \sum_{k=0}^{\frac{d-1}{2}} \beta_{k,d+1} (-1)^k \sin \left(  \frac{\pi s}{2} \right) m^{2k+1-s} \Gamma (2k+1) \prod_{i=1}^{2k+1} \frac{1}{s-i} \,.
\end{aligned}
    \label{zeta_hyperbolic_odd}
\end{align}
We immediately obtain 
\begin{align}
    \zeta_{\mathbb{H}^d} (0,m)& = 0 \,, \\
\partial_s \zeta_{\mathbb{H}^d} (0,m)&= c_d \, r_d \sum_{k=0}^{\frac{d-1}{2}} \beta_{k,d+1} \frac{\pi m^{2k+1} }{2k+1} (-1)^{k+1} \,.
\end{align}
For the massive fermion, the renormalized free energy is given by 
\begin{align} \label{FE_ren_odd}
\begin{aligned}
    F_\text{ren} [\mathbb{H}^d] (m) & =  \frac{1}{2} \partial_s \zeta_{\mathbb{H}^d} (0,m) \\
    & =\frac{ (-2)^{\frac{d-1}{2}}}{\Gamma(d)} \sum_{k=0}^{\frac{d-1}{2}} \beta_{k,d+1} \frac{m^{2k+1} }{2k+1} (-1)^{k+1}  \log \left( \frac{R}{\epsilon} \right) \,.
\end{aligned}
\end{align}
In the massless limit,  the renormalized free energy vanishes,
\begin{align} \label{hyperbolic_FE_odd}
    F_\text{ren} [\mathbb{H}^d] = 0 \,.
\end{align}
Here and hereafter, we omit $(0)$ in free energies for the massless fermion.
Equation \eqref{hyperbolic_FE_odd} implies that the boundary anomaly does not exist.

\subsubsection{Even \texorpdfstring{$d$}{d}}
\label{sec4.1.2}

Using the expansion of the Plancherel \eqref{Plancherel}
\begin{align} \label{Plancherel_even}
    \mu^{(d)} (\omega) 
    = c_d \, r_d \coth (\pi \omega) \sum_{k=1}^{\frac{d}{2}} \alpha_{k,d+1} \omega^{2k-1} \,,
\end{align}
and the identity
\begin{align} \label{coth}
    \coth (\pi \omega) = 1+ \frac{2}{\mathrm{e}^{2\pi \omega}-1} \,,
\end{align}
the zeta function can be decomposed into two parts,
\begin{align}
    \zeta_{\mathbb{H}^d} (s,m) &= \zeta_{\mathbb{H}^d}^{(1)} (s,m) + \zeta_{\mathbb{H}^d}^{(2)} (s,m) \,, \\
    \zeta_{\mathbb{H}^d}^{(1)} (s,m) & = c_d \, r_d  \sum_{k=1}^{\frac{d}{2}} \alpha_{k,d+1} \int_0^\infty \! \dd \omega \, \omega^{2k-1} \left( (\omega +\i \, m)^{-s} + (\omega -\i \, m)^{-s} \right) \,, \\
    \zeta_{\mathbb{H}^d}^{(2)} (s,m) & =
    2c_d \, r_d \sum_{k=1}^{\frac{d}{2}} \alpha_{k,d+1} \int_0^\infty \! \dd \omega \,  \frac{\omega^{2k-1}}{\mathrm{e}^{2\pi \omega}-1}  \, \left((\omega +\i \, m)^{-s} + (\omega -\i \, m)^{-s} \right) \,.
\end{align}
The first term of the zeta function can be calculated as 
\begin{align}  
    \zeta_{\mathbb{H}^d}^{(1)} (s,m) = 2c_d \, r_d  \sum_{k=1}^{\frac{d}{2}} \alpha_{k,d+1} m^{2k-s} (-1)^k \cos  \left( \frac{ \pi s}{2}\right) \Gamma (2k) \prod_{i=1}^{2k} \frac{1}{s-i} \,,
\end{align}
and we can easily read off 
\begin{align}  
    \zeta_{\mathbb{H}^d}^{(1)} (0,m) & = c_d \, r_d \sum_{k=1}^{\frac{d}{2}} \alpha_{k,d+1} \frac{(-1)^k}{k} m^{2k} \,, \\
    \partial_s \zeta_{\mathbb{H}^d}^{(1)} (0,m) & = 
    c_d \, r_d \sum_{k=1}^{\frac{d}{2}} \alpha_{k,d+1} \frac{(-1)^k}{k} m^{2k} \left( H_{2k} -\log m \right) \,, 
\end{align}
where $H_{2k}$ is a harmonic number.

On the other hand, it is difficult to perform the integral in  $\zeta_{\mathbb{H}^d}^{(2)} (s,m)$, hence we compute $\zeta_{\mathbb{H}^d}^{(2)} (0,m)$ and $\partial_s \zeta_{\mathbb{H}^d}^{(2)} (0,m)$ instead.
$\zeta_{\mathbb{H}^d}^{(2)} (0,m)$ is independent on the mass term, 
\begin{align}  
    \zeta_{\mathbb{H}^d}^{(2)} (0,m) 
    =c_d \, r_d \sum_{k=1}^{\frac{d}{2}} \alpha_{k,d+1} (-1)^{k+1} \frac{B_{2k}}{k}
    \,,
\end{align}
and the derivative can be computed as 
\begin{align} \label{dzeta2}
    \partial_s \zeta_{\mathbb{H}^d}^{(2)} (0,m) = - 2c_d \, r_d \sum_{k=1}^{\frac{d}{2}} \alpha_{k,d+1} f_k (m)
\end{align}
with 
\begin{align} \label{fkm}
\begin{aligned}
    f_k(m)  
& = \int_0^\infty \! \dd \omega \,  \frac{\omega^{2k-1}}{\mathrm{e}^{2\pi \omega}-1} \log (\omega^2 + m^2 ) \\
& =
    (-1)^{k}  \left[  \frac{m^{2k-1}}{2(2k-1)} + \frac{m^{2k}}{4k} \left( \frac{1}{k} - \log( m^2) \right)  + \sum_{l=1}^{k-1} \frac{B_{2l}}{4l} \frac{m^{2k-2l}}{k-l} - \frac{B_{2k}H_{2k-1}}{2k} \right. \\
    & \left. \phantom{(-1)^{k}} \qquad  +  \sum_{r=0}^{2k-1} (-1)^r \binom{2k-1}{r} m^{2k-1-r} \left( \zeta'(-r,m) -  \frac{B_{r+1}(m) H_r}{r+1} \right) \right] \,,
\end{aligned}
\end{align}
where $B_{r+1} (m)$ is a Bernoulli polynomial.
See appendix \ref{app2} for the detailed derivation of \eqref{fkm}.

We obtain the renormalized free energy for the massive fermion,
\begin{align}
    F_{\text{ren}}[\mathbb{H}^d] (m) = - A[\mathbb{H}^d] (m) \log (\Lambda R) +F_\text{fin} [\mathbb{H}^d] (m)
\end{align}
with the coefficient of the logarithmic divergent part
\begin{align}
\begin{aligned} \label{anomaly_even}
    A[\mathbb{H}^d] (m) & = - \frac{1}{2} \left( \zeta_{\mathbb{H}^d}^{(1)} (0,m)+\zeta_{\mathbb{H}^d}^{(2)} (0,m) \right) \\
    & = c_d \, r_d \sum_{k=1}^{\frac{d}{2}} \alpha_{k,d+1} (-1)^{k} \frac{B_{2k}-m^{2k}}{2k}
     \,,
\end{aligned}
\end{align}
and the finite part
\begin{align} \label{finite_even}
\begin{aligned}
    & F _\text{fin} [\mathbb{H}^d] (m)  =  \frac{1}{2} \left( \partial_s \zeta_{\mathbb{H}^d}^{(1)} (0,m) + \partial_s \zeta_{\mathbb{H}^d}^{(2)} (0,m) \right) \\
    & \qquad = c_d \, r_d \sum_{k=1}^{\frac{d}{2}} \alpha_{k,d+1} (-1)^{k+1}\left( - m^{2k} \frac{H_{2k-1}}{2k} -\frac{B_{2k}H_{2k-1}}{2k} +  \frac{m^{2k-1}}{2(2k-1)} \right.  \\
    & \left. \qquad  \quad + \sum_{l=1}^{k-1} \frac{B_{2l}}{4l} \frac{m^{2k-2l}}{k-l} +  \sum_{r=0}^{2k-1} (-1)^r \binom{2k-1}{r} m^{2k-1-r} \left( \zeta'(-r,m) -  \frac{B_{r+1}(m) H_r}{r+1} \right) \right) \,.
\end{aligned}
\end{align}
In the massless limit, $m \to 0 $, the renormalized free energy has a simple expression
\begin{align}
    F_{\text{ren}}[\mathbb{H}^d] = - A[\mathbb{H}^d] \log (\Lambda R) +F_\text{fin} [\mathbb{H}^d] \,, 
\end{align}
with the anomaly coefficient and the finite part
\begin{align}
    A[\mathbb{H}^d] &= c_d \, r_d \sum_{k=1}^{\frac{d}{2}} \alpha_{k,d+1} (-1)^{k+1} \frac{B_{2k}}{2k} \,, \\
    F _\text{fin} [\mathbb{H}^d] 
    & = c_d \, r_d \sum_{k=1}^{\frac{d}{2}} \alpha_{k,d+1} (-1)^k  \zeta'(1-2k) \,.
\end{align}
To see the massless limit of the finite term, it is convenient to use \eqref{b.7} instead of \eqref{fkm}.
The free energy on $\mathbb{H}^d$ is just the half of that on $\mathbb{S}^d$, 
\begin{align}
    F_\text{ren}[\mathbb{H}^d] = \frac{1}{2}F_\text{ren}[\mathbb{S}^d] \,,
\end{align}
and this reproduces the known anomaly coefficients in literature \cite{Rodriguez-Gomez:2017aca}.

The anomaly coefficients and the finite parts of the free energy for $d \leq 10$ are listed in tables \ref{tab:F_p_q} and \ref{tab:Fin_p_q} in appendix \ref{app_table}.
 
\subsection{Free energy on \texorpdfstring{$\mathbb{H}^{p+1}\times \mathbb{S}^{q-1}$}{Hp-1,Sq-1}}
\label{sec4.2}

The free energy on $\mathbb{H}^{p+1} \times \mathbb{S}^{q-1}$ for the massless Dirac fermion except for even $p$ and even $q \geq 4$ is expressed as 
\begin{align}
    F[\mathbb{H}^{p+1} \times \mathbb{S}^{q-1}] = -
    \frac{1}{2}
    \sum_{\ell=0}^{\infty} g^{(q-1)}(\ell) \int_0^\infty \! \dd \omega \, \mu^{(p+1)} (\omega )  \log \left( \frac{\omega^2 +\left(\nu_\ell^{(q-1)}\right)^2}{\tilde{\Lambda}^2 R^2} \right) \,, 
\end{align}
with the Plancherel measure \eqref{Plancherel} and the degeneracy \eqref{degeneracy}.
For even $p$ and even $q \geq 4$, there are two fermions of opposite $U(1)$ charge as the additional $U(1)$ appeared in \eqref{field_decomposition}.
Then, the free energy on $\mathbb{H}^{p+1} \times \mathbb{S}^{q-1}$ with even $p$ and even $q \geq 4$ becomes 
\begin{align}
    F[\mathbb{H}^{p+1} \times \mathbb{S}^{q-1}] = -
    \sum_{\ell=0}^{\infty} g^{(q-1)}(\ell) \int_0^\infty \! \dd \omega \, \mu^{(p+1)} (\omega )  \log \left( \frac{\omega^2 +\left(\nu_\ell^{(q-1)}\right)^2}{\tilde{\Lambda}^2 R^2} \right) \,.
\end{align}
To treat these two cases simultaneously, we introduce a notation,
\begin{align}
     d_{U(1)}= 
     \begin{dcases}
       2 & \text{for even } p \text{ and even } q \geq 4\,,  \\
       1 & \text{otherwise} \,.
     \end{dcases}
\end{align}

In the following, we compute the free energy using the zeta-function regularisation divided into two cases: even $p$ case and odd $p$ case.
An important notice is that we use different decompositions for the logarithmic function depending on the evenness of $p$.

\subsubsection{Even \texorpdfstring{$p$}{p}}

Performing similar computations in section \ref{sec4.1.1}, the renormalized free energy is given by
\begin{align}
    F_\text{ren} [\mathbb{H}^{p+1} \times \mathbb{S}^{q-1}]  =  \frac{1}{2} \zeta_{\mathbb{H}^{p+1} \times \mathbb{S}^{q-1}} (0) \log(\Lambda R) + \frac{1}{2} \partial_s \zeta_{\mathbb{H}^{p+1} \times \mathbb{S}^{q-1}}(0)
\end{align}
with the zeta function
\begin{align}
    \zeta_{\mathbb{H}^{p+1} \times \mathbb{S}^{q-1}} (s) = d_{U(1)}\sum_{\ell=0}^\infty g^{(q-1)} (\ell) \zeta_{\mathbb{H}^{p+1}} \left(s,\nu_\ell^{(q-1)} \right)  \,.
\end{align}
Here $\zeta_{\mathbb{H}^{p+1}} \left(s,\nu_\ell^{(q-1)} \right)$ is the zeta function of $\mathbb{H}^d$  with mass $\nu_\ell^{(q-1)}$ \eqref{zeta_hyperbolic_odd}.
By using  the expansion of the degeneracy \eqref{expansion} with the replacement $d \to q-1$, the zeta function becomes 
\begin{align}
\begin{aligned}
    \zeta_{\mathbb{H}^{p+1} \times \mathbb{S}^{q-1}} (s) 
    &     = 2d_{U(1)}c_{p+1} \,  r_{p+1}\sum_{k=0}^{\frac{p}{2}} \beta_{k,p+2} (-1)^k  \sin \left( \frac{\pi s}{2} \right)  \Gamma (2k+1) \prod_{i=1}^{2k+1} \frac{1}{s-i} \\
    & \quad \cdot  \frac{2r_{q-1}}{\Gamma (q-1)} 
    \begin{dcases}
        \sum_{n=0}^{\frac{q-1}{2}}(-1)^{\frac{q-1}{2}+n}\,\alpha_{n, q}\,
        \zeta ( s-2n-2k)
        & q: \text{odd} \,, \\
         \sum_{n=0}^{\frac{q-2}{2}}(-1)^{\frac{q}{2}-1+n}\,\beta_{n, q}\, \zeta_\text{H} \left( s-2n-2k-1,\frac{1}{2} \right) 
         & q: \text{even} \,. 
    \end{dcases}
\end{aligned}
\end{align}
We immediately find 
\begin{align}
    \zeta_{\mathbb{H}^{p+1} \times \mathbb{S}^{q-1}} (0)  = 0 \,,
\end{align}
and 
\begin{align}
\begin{aligned}
    \partial_s \zeta_{\mathbb{H}^{p+1} \times \mathbb{S}^{q-1}} (0) 
    &= -d_{U(1)}c_{p+1} \, r_{p+1} \sum_{k=0}^{\frac{p}{2}} \beta_{k,p+2} \frac{(-1)^k \pi}{2k+1}   \\
    & \cdot  \frac{2r_{q-1}}{\Gamma (q-1)} 
    \begin{dcases}
        0 & q: \text{odd} \,, \\
         \sum_{n=0}^{\frac{q-2}{2}}(-1)^{\frac{q}{2}-1+n}\,\beta_{n, q}\, \zeta_\text{H} \left( -2n-2k-1,\frac{1}{2} \right) 
         & q: \text{even} \,.
    \end{dcases}
\end{aligned}
\end{align}

For even $p$, we find the following:
\begin{itemize}
    \item For odd $q$, the renormalized free energy vanishes,
    \begin{align} \label{relation0}
        F_\text{ren}[\mathbb{H}^{p+1} \times \mathbb{S}^{q-1}] =0 \,.
    \end{align}
    This implies that both bulk and defect anomalies vanish.
    \item For even $q$, the renormalized free energy has a defect anomaly which comes from the volume of $\mathbb{H}^{p+1}$, 
     \begin{align}
        F_\text{ren}[\mathbb{H}^{p+1} \times \mathbb{S}^{q-1}] = - \mathcal{A} [\mathbb{H}^{p+1} \times \mathbb{S}^{q-1}] \log \left( \frac{R}{\epsilon} \right)
    \end{align}
    with 
    \begin{align}
    \begin{aligned}
         \mathcal{A} [\mathbb{H}^{p+1} \times \mathbb{S}^{q-1}]&= - 
         \frac{ (-2)^\frac{p+q}{2}d_{U(1)}}{\Gamma(p+1)\Gamma (q-1)} \,  \sum_{k=0}^{\frac{p}{2}} \beta_{k,p+2} \frac{(-1)^k}{2k+1}   
         \\
    & \qquad \cdot 
         \sum_{n=0}^{\frac{q}{2}-1}(-1)^{n}\,\beta_{n, q}\, \zeta_\text{H} \left( -2n-2k-1,\frac{1}{2} \right) \,.
    \end{aligned}
    \end{align}
    \item For $q=p+2$, the defect anomaly is proportional to the bulk anomaly on $\mathbb{S}^{2p+2}$, 
    \begin{align} \label{anomaly_relation}
        A[\mathbb{S}^{2p+2}] = 2 \mathcal{A} [\mathbb{H}^{p+1} \times \mathbb{S}^{p+1}] \,,
    \end{align}
    as in the scalar case \cite{Rodriguez-Gomez:2017kxf,Nishioka:2021uef} and the holographic case \cite{Rodriguez-Gomez:2017kxf}.
    It is expected that the holographic result should be the same as the free fermion result 
because the anomaly coefficient does not depend on the strength of a coupling constant and the holography can be applied if the number of fermions is large.
\end{itemize}
For $q=2$, our result correctly reproduces the free energy obtained in \cite{Klebanov:2011uf,Beccaria:2017dmw} (with an appropriate dimension of the spinor).

The anomaly coefficients of the free energy are listed in table \ref{tab:F_p_q} in appendix \ref{app_table}.

\subsubsection{Odd \texorpdfstring{$p$}{p}}

In this section, we do not omit all equations in order to derive our main results of this section, \eqref{relation1} and \eqref{466}. If the reader is not interested in the detail of the derivations, the reader can skip until \eqref{relation1}.

For odd $p$, the Plancherel measure \eqref{Plancherel_even} is decomposed into two parts using the identity \eqref{coth}, and the free energy consists of two parts,
\begin{align} \label{FE_even_p_odd_q_def}
\begin{aligned}
     F[\mathbb{H}^{p+1} \times \mathbb{S}^{q-1}]
    & = - \frac{c_{p+1}\, r_{p+1}}{2} \sum_{k=1}^{\frac{p+1}{2}} \alpha_{k,p+2} \sum_{\ell=0}^{\infty} g^{(q-1)}(\ell)
    \int_0^\infty \! \dd \omega \, \omega^{2k-1}  
    \\
    & \qquad \qquad \cdot 
    \left[ \log \left( \frac{\omega + \i \, \nu_\ell^{(q-1)}}{\tilde{\Lambda} R} \right) + \log \left( \frac{\omega -\i \,  \nu_\ell^{(q-1)}}{\tilde{\Lambda} R} \right)\right] \\
    & \quad \phantom{=} - c_{p+1}\,  r_{p+1} \sum_{k=1}^{\frac{p+1}{2}} \alpha_{k,p+2} 
     \int_0^\infty \! \dd \omega \,\frac{\omega^{2k-1}}{\mathrm{e}^{2\pi \omega}-1} \sum_{\ell=0}^{\infty} g^{(q-1)}(\ell) 
     \\
    & \qquad \qquad \cdot 
    \left[ \log \left( \frac{\nu_\ell^{(q-1)}+ \i \, \omega  }{\tilde{\Lambda} R} \right) + \log \left( \frac{\nu_\ell^{(q-1)} -\i \,  \omega}{\tilde{\Lambda} R} \right)\right] \,.
\end{aligned}
\end{align}
In contrast to section \ref{sec4.1.2}, we decompose the logarithmic function differently for each term, and the ordering of the summation over $\ell$ and the integral over $\omega$ is important.

The Schwinger representation of the free energy \eqref{FE_even_p_odd_q_def} is given by
\begin{align}
\begin{aligned}
     F[\mathbb{H}^{p+1} \times \mathbb{S}^{q-1}] 
    & =  \frac{c_{p+1}\, r_{p+1} }{2} \sum_{k=1}^{\frac{p+1}{2}} \alpha_{k,p+2} \int_0^\infty \! \frac{\dd t}{t} \sum_{\ell=0}^{\infty} g^{(q-1)}(\ell)
    \int_0^\infty \! \dd \omega \,  \omega^{2k-1} \\
     & \qquad \qquad \cdot 
    \left[ \mathrm{e}^{-t(\omega + \i \,  \nu_\ell^{(q-1)})/(\tilde{\Lambda} R) } + \mathrm{e}^{-t(\omega - \i \,  \nu_\ell^{(q-1)})/(\tilde{\Lambda} R) } \right] \\
    & \quad \phantom{=} + c_{p+1} \, r_{p+1} \sum_{k=1}^{\frac{p+1}{2}} \alpha_{k,p+2} \int_0^\infty \! \frac{\dd t}{t}
     \int_0^\infty \! \dd \omega \,  \frac{\omega^{2k-1}}{\mathrm{e}^{2\pi \omega}-1} \sum_{\ell=0}^{\infty} g^{(q-1)}(\ell) \\
     & \qquad \qquad \cdot 
    \left[ \mathrm{e}^{-t(\nu_\ell^{(q-1)}+ \i \,  \omega )/(\tilde{\Lambda} R) } + \mathrm{e}^{-t(\nu_\ell^{(q-1)} - \i \, \omega)/(\tilde{\Lambda} R) } \right] \,.
\end{aligned}
\end{align}
For the first term, we first perform the integral over $\omega$, and hence the first term is convergent in the $\omega \to \infty$ limit in the Schwinger representation. 
On the other hand, for the second term, we first perform the summation over $\ell$, and hence the second term is convergent in the $\ell \to \infty$ limit in the Schwinger representation. 

The renormalized free energy is given by 
\begin{align}
    F_\text{ren} [\mathbb{H}^{p+1} \times \mathbb{S}^{q-1}]  = \frac{1}{2} \zeta_{\mathbb{H}^{p+1} \times \mathbb{S}^{q-1}} (0) \log(\Lambda R) + \frac{1}{2} \partial_s \zeta_{\mathbb{H}^{p+1} \times \mathbb{S}^{q-1}}(0) \,, 
\end{align}
where the zeta function is a sum of two parts, 
\begin{align}
    \zeta_{\mathbb{H}^{p+1} \times \mathbb{S}^{q-1}}(s) = \zeta_{\mathbb{H}^{p+1} \times \mathbb{S}^{q-1}}^{(1)} (s) + \zeta_{\mathbb{H}^{p+1} \times \mathbb{S}^{q-1}}^{(2)} (s) \,, 
\end{align}  
with 
\begin{align}
&
\begin{aligned}
     \zeta_{\mathbb{H}^{p+1} \times \mathbb{S}^{q-1}}^{(1)} (s) & =  c_{p+1} \, r_{p+1} \sum_{k=1}^{\frac{p+1}{2}} \alpha_{k,p+2} \sum_{\ell=0}^{\infty} g^{(q-1)}(\ell)    \\
    & \qquad \cdot \int_0^\infty \! \dd \omega \,  \omega^{2k-1} 
    \left[ \left( \omega +\i \, \nu_\ell^{(q-1)} \right)^{-s} + \left( \omega -\i \, \nu_\ell^{(q-1)} \right)^{-s} \right] \,, 
\end{aligned}
    \\
   & \begin{aligned}
    \zeta_{\mathbb{H}^{p+1} \times \mathbb{S}^{q-1}}^{(2)} (s) & =
    2 c_{p+1} \, r_{p+1} \sum_{k=1}^{\frac{p+1}{2}} \alpha_{k,p+2} 
     \int_0^\infty \! \dd \omega \,  \frac{\omega^{2k-1}}{\mathrm{e}^{2\pi \omega}-1} 
    \\
    & \qquad \cdot \sum_{\ell=0}^{\infty} g^{(q-1)}(\ell)
    \left[ \left( \nu_\ell^{(q-1)}+ \i \, \omega \right)^{-s} +  \left( \nu_\ell^{(q-1)} -\i \, \omega \right)^{-s} \right] \,.       
    \end{aligned}
\end{align}  
The first term in the zeta function is convergent for $\text{Re}\, s> 2k$,
\begin{align}
\begin{aligned}
    \zeta_{\mathbb{H}^{p+1} \times \mathbb{S}^{q-1}}^{(1)} (s) 
    & = 2 c_{p+1} r_{p+1} \sum_{k=1}^{\frac{p+1}{2}} \alpha_{k,p+2}  (-1)^k \cos \left( \frac{\pi s}{2} \right) \Gamma (2k) \prod_{i=1}^{2k} \frac{1}{s-i} \\ 
    & \quad \cdot  \frac{2r_{q-1} }{\Gamma (q-1)} \begin{dcases}
        \sum_{n=0}^{\frac{q-1}{2}}(-1)^{\frac{q-1}{2}+n}\,\alpha_{n, q}\, \zeta ( s-2n-2k+1 ) & q: \text{odd} \,,\\
         \sum_{n=0}^{\frac{q}{2}-1}(-1)^{\frac{q}{2}-1+n}\,\beta_{n, q}\,\zeta_\text{H} \left( s-2n-2k,\frac{1}{2} \right) & q: \text{even}\,.
    \end{dcases}
\end{aligned}
\end{align}  
It follows that 
\begin{align}
\begin{aligned}
    \zeta_{\mathbb{H}^{p+1} \times \mathbb{S}^{q-1}}^{(1)} (0) 
    & = c_{p+1} \, r_{p+1} \frac{2r_{q-1}}{\Gamma (q-1)} \sum_{k=1}^{\frac{p+1}{2}} \alpha_{k,p+2}  \frac{(-1)^k}{k}  
    \\ 
    & \quad \cdot  
    \begin{dcases}
        \sum_{n=0}^{\frac{q-1}{2}}(-1)^{\frac{q-1}{2}+n}\,\alpha_{n, q}\, \zeta ( -2n-2k+1 ) & q: \text{odd}\,, \\
         0  & q: \text{even} \,, 
    \end{dcases}
\end{aligned}
\end{align} 
and 
\begin{align}
\begin{aligned}
    & \partial_s \zeta_{\mathbb{H}^{p+1} \times \mathbb{S}^{q-1}}^{(1)} (0) \\
    & \quad =  c_{p+1} \, r_{p+1} \frac{2r_{q-1}}{\Gamma (q-1)}  \sum_{k=1}^{\frac{p+1}{2}} \alpha_{k,p+2} \frac{(-1)^k}{k} \\ 
    & \qquad \cdot \begin{dcases}
        \sum_{n=0}^{\frac{q-1}{2}}(-1)^{\frac{q-1}{2}+n}\,\alpha_{n, q}\, \left[ H_{2k} \zeta ( -2n-2k+1 )+ \zeta ' ( -2n-2k+1 ) \right]& q: \text{odd} \,, \\
         \sum_{n=0}^{\frac{q}{2}-1}(-1)^{\frac{q}{2}-1+n}\,\beta_{n, q}\,
         \partial_s \zeta_\text{H} \left( -2n-2k,\frac{1}{2} \right) 
         & q: \text{even} \,.
    \end{dcases}
\end{aligned}
\end{align} 
Using the expansion of the gamma functions in the degeneracy,
\begin{align} 
\begin{aligned}
    & \frac{\Gamma \left(\nu_\ell^{(q-1)} +\frac{q-1}{2} \right) }{ \Gamma \left(\nu_\ell^{(q-1)} - \frac{q-1}{2}+1 \right)} 
    \\
    & \qquad  = \begin{dcases}
        \sum_{n=0}^{\frac{q-1}{2}}(-1)^{\frac{q-1}{2}+n}\,\alpha_{n, q}\, \sum_{i=0}^{2n-1} \binom{2n-1}{i} \left(\nu_\ell^{(q-1)} + \i \,  \omega \right)^{i} (-\i \,  \omega)^{2n-1-i} & q: \text{odd} \,, \\
         \sum_{n=0}^{\frac{q}{2}-1}(-1)^{\frac{q}{2}-1+n}\,\beta_{n, q}\, \sum_{i=0}^{2n} \binom{2n}{i}  \left(\nu_\ell^{(q-1)}+\i \,  \omega \right)^{i} (-\i \,  \omega)^{2n-i} & q: \text{even} \,,
    \end{dcases}
\end{aligned}
\end{align}
and its complex conjugate, we obtain
\begin{align}
\begin{aligned}
    \zeta_{\mathbb{H}^{p+1} \times \mathbb{S}^{q-1}}^{(2)} (s) 
    & =2 c_{p+1} \, r_{p+1} \sum_{k=1}^{\frac{p+1}{2}} \alpha_{k,p+2} 
     \int_0^\infty \! \dd \omega \,  \frac{\omega^{2k-1}}{\mathrm{e}^{2\pi \omega}-1} \frac{2r_{q-1}}{\Gamma (q-1)} \\
     & \qquad \cdot 
     \begin{dcases}
     \begin{aligned}
          & \sum_{n=0}^{\frac{q-1}{2}}(-1)^{\frac{q-1}{2}+n}\,\alpha_{n, q}\, \sum_{i=0}^{2n-1} \binom{2n-1}{i} (-\i)^{2n-1-i} \omega^{2n-1-i} \\
        & \quad \cdot \left[ \zeta_\text{H} 
        \left(s-i, \i \, \omega \right)  + (-1)^{i+1} \zeta_\text{H} 
        \left(s-i, - \i \, \omega \right) \right]
     \end{aligned}
         & q: \text{odd} \,, \\
        \begin{aligned}
         &\sum_{n=0}^{\frac{q}{2}-1}(-1)^{\frac{q}{2}-1+n}\,\beta_{n, q}\, \sum_{i=0}^{2n} \binom{2n}{i} (-\i)^{2n-i} \omega^{2n-i} \\ 
         & \quad \cdot \left[ \zeta_\text{H}  \left(s-i, \frac{1}{2}+\i \, \omega \right) + (-1)^i \zeta_\text{H} \left(s-i, \frac{1}{2}-\i \, \omega \right)  \right]  
        \end{aligned}
         & q: \text{even} \,.
    \end{dcases} 
\end{aligned}
\end{align}  
Although it is difficult to perform the integral over $\omega$ analytically, it is possible to simplify $\zeta_{\mathbb{H}^{p+1} \times \mathbb{S}^{q-1}}^{(2)} (0)$ and its derivative at $s=0$ furthermore.
After a bit of calculation, we obtain
\begin{align}
\begin{aligned}
    \zeta_{\mathbb{H}^{p+1} \times \mathbb{S}^{q-1}}^{(2)} (0) =
    \begin{dcases}
      \begin{aligned}
           & c_{p+1} \, r_{p+1} \frac{2r_{q-1}}{\Gamma (q-1)} \sum_{k=1}^{\frac{p+1}{2}} \alpha_{k,p+2} 
        \sum_{n=0}^{\frac{q-1}{2}}(-1)^{\frac{q-1}{2}}\,\alpha_{n, q}
        \\
     & \quad \cdot \sum_{i=0}^{2n-1} \binom{2n-1}{i} \frac{(-1)^{i+k+n+1}}{i+1} \sum_{l=0}^{\floor{\frac{i+1}{2}}} \binom{i+1}{2l} B_{2l} \frac{B_{2k+2n-2l}}{k+n-l} 
      \end{aligned} & q: \text{odd} \,, \\
       0 & q: \text{even} \,.
    \end{dcases}
\end{aligned}
\end{align} 
In addition, the derivative of the zeta function at $s=0$ can be computed as
\begin{align}\label{integral}
\begin{aligned}
    \partial_s \zeta_{\mathbb{H}^{p+1} \times \mathbb{S}^{q-1}}^{(2)} (0) 
    & =2 c_{p+1} \, r_{p+1} \sum_{k=1}^{\frac{p+1}{2}} \alpha_{k,p+2} 
     \int_0^\infty \! \dd \omega \,  \frac{\omega^{2k-1}}{\mathrm{e}^{2\pi \omega}-1} \frac{2r_{q-1}}{\Gamma (q-1)} \\
     & \qquad \cdot 
     \begin{dcases}
       \begin{aligned}
         & \sum_{n=0}^{\frac{q-1}{2}}(-1)^{\frac{q-1}{2}+n}\,\alpha_{n, q}\, \sum_{i=0}^{2n-1} \binom{2n-1}{i} (-\i)^{2n-1-i} \omega^{2n-1-i} \\
        & \quad \cdot \left[ \partial_s \zeta_\text{H} 
        \left(-i, \i \, \omega \right)  + (-1)^{i+1} \partial_s \zeta_\text{H} 
        \left(-i, - \i \, \omega \right) \right]   
        \end{aligned}
        & q: \text{odd} \,, \\
         \begin{aligned}
          &\sum_{n=0}^{\frac{q}{2}-1}(-1)^{\frac{q}{2}-1+n}\,\beta_{n, q}\, \sum_{i=0}^{2n} \binom{2n}{i} (-\i)^{2n-i} \omega^{2n-i} \\ 
         &\quad \cdot \left[ \partial_s \zeta_\text{H}  \left(-i, \frac{1}{2}+\i \, \omega \right) + (-1)^i \partial_s \zeta_\text{H} \left(-i, \frac{1}{2}-\i \, \omega \right)  \right]     
         \end{aligned}
          & q: \text{even} \,.
    \end{dcases} 
\end{aligned}
\end{align}  
For even $q$, the combination of the Hurwitz zeta functions can be simplified using a formula
\begin{align}
\begin{aligned}
    & \partial_s \zeta_\text{H}  \left(-i, \frac{1}{2}+\i \, \omega \right) + (-1)^i \partial_s \zeta_\text{H} \left(-i, \frac{1}{2}-\i \, \omega \right) \\
    & \qquad = \frac{\Gamma (i+1)}{(2\pi \i)^i} \mathrm{Li}_{i+1} (-\mathrm{e}^{-2\pi \omega}) +  \frac{\pi \, \i}{i+1}B_{i+1} \left( \frac{1}{2} + \i \, \omega \right) \,.
\end{aligned}
\end{align}
See e.g. (B.19) in \cite{Nishioka:2021uef} for the derivation.
Unfortunately, it is difficult to simplify the equations anymore, so we perform the integral \eqref{integral} numerically.

For odd $p$, we find the following:
\begin{itemize}
    \item For odd $q$, we numerically find the relation among free energies 
    \begin{align} \label{relation1}
        F_\text{ren}[\mathbb{S}^d]  = F_\text{ren}[\mathbb{H}^{2k} \times \mathbb{S}^{d-2k}] 
    \end{align}
    for $k=1,2,\cdots , d/2-1$.
    That is, the anomaly coefficients and the finite parts satisfy the relations,
    \begin{align} \label{relation2}
        A[\mathbb{S}^d] & = A[\mathbb{H}^{2k} \times \mathbb{S}^{d-2k}] \,, \\
        F_\text{fin}[\mathbb{S}^d] & = F_\text{fin}[\mathbb{H}^{2k} \times \mathbb{S}^{d-2k}] \,.
        \label{relation3}
    \end{align}
    \item For even $q$, the anomaly parts vanish since the bulk dimension $d=p+q$ is odd.
    We numerically find the relation between the universal parts of the free energy 
    \begin{align} \label{466}
        F_\text{fin}[\mathbb{S}^d] = F_\text{fin}[\mathbb{H}^{2k} \times \mathbb{S}^{d-2k} ] 
    \end{align}
    for $k=1,2,\cdots, (d-1)/2$.
\end{itemize}
The equivalence of the anomaly coefficients  between $\mathbb{H}^{2k} \times \mathbb{S}^{d-2k}$ and $\mathbb{S}^d$ follows from a relation of Euler characteristic $\chi [\mathbb{H}^{2k} \times \mathbb{S}^{d-2k}] = \chi [\mathbb{S}^d]$ as pointed out \cite{Rodriguez-Gomez:2017kxf,Nishioka:2021uef} because the bulk anomaly is related to the Euler characteristic.

\section{Free energy for Neumann boundary condition}
\label{sec5}

In section \ref{sec4}, we computed the free energy on $\mathbb{H}^{p+1}\times \mathbb{S}^{q-1}$ with Dirichlet boundary condition.
In this section, we compute the free energy with Neumann boundary condition.
Neumann boundary condition exists only when $q=2$ as we saw in section \ref{sec2}.

\subsection{Analytical continuation}
\label{sec5.1}

We decompose the free energy into a sum of $\ell$,
\begin{align}
    F [\mathbb{H}^{p+1} \times \mathbb{S}^{q-1}]= \sum_{\ell=0}^\infty g^{(q-1)} (\ell) F_\ell \left(\nu_\ell^{(q-1)} \right) \,, 
\end{align}
where $F_\ell \left(\nu_\ell^{(q-1)} \right)$ is the free energy for the $\ell$-th mode on $\mathbb{H}^{p+1}$,
\begin{align}
    F_\ell \left(\nu_\ell^{(q-1)} \right) =  -\frac{d_{U(1)}}{2}  \int_0^\infty \dd \omega \, \mu^{(p+1)} (\omega)\,  \left[ \log \left( \frac{\omega + \i \, \nu_\ell^{(q-1)}}{\tilde \Lambda R}  \right) + \log \left( \frac{\omega - \i \, \nu_\ell^{(q-1)}}{\tilde \Lambda R}  \right) \right] \,.
\end{align}
Since the $\omega$ integral is   performed before the summation over $\ell$, the logarithmic function is decomposed in this way.

From now on, we concentrate on $q=2$ because Neumann boundary condition is allowed only when $q=2$.
The difference of free energies between the two boundary conditions comes from the $\ell=0$ mode.
The Dirichlet boundary condition has a positive value
\begin{align}
    \nu_{\ell=0}^{(1)} =  \frac{1}{2} \,,
\end{align}
while the Neumann boundary condition has a negative value
\begin{align}
    \nu_{\ell=0}^{(1)} = - \frac{1}{2} \,.
\end{align}
In section \ref{sec4.1}, we obtained the zeta functions as analytical functions of positive $m$. 
It is possible to analytically continue the zeta functions to a $m<0$ region.

\paragraph{Even $p$}

From \eqref{FE_ren_odd} the free energy of the $\ell=0$ mode is given by
\begin{align} 
    F_{\ell =0} \left(\nu_{\ell=0}^{(1)} \right)  
    =\frac{(-1)^{\frac{p}{2}+1} 2^{\frac{p}{2}}}{\Gamma(p+1)} \sum_{k=0}^{\frac{p}{2}} \beta_{k,p+2} \frac{\left(\nu_{\ell=0}^{(1)} \right)^{2k+1} }{2k+1} (-1)^{k}  \log \left( \frac{R}{\epsilon} \right) \,.
\end{align}
Then, the difference of the free energies becomes 
\begin{align}
    F_{\Delta_\text{D}}[\mathbb{H}^{p+1} \times \mathbb{S}^1] - F_{\Delta_\text{N}}[\mathbb{H}^{p+1} \times \mathbb{S}^1] = \frac{(-2)^{\frac{p}{2}+1}}{\Gamma(p+1)} \sum_{k=0}^{\frac{p}{2}} \beta_{k,p+2} \frac{(-1)^{k}}{2^{2k}(2k+1)}  \log \left( \frac{R}{\epsilon} \right) \,.
\end{align}
By using 
\begin{align}
    \frac{1}{2^{2k}(2k+1)} = 2  \int_0^{\frac{1}{2}} \! \dd u \, u^{2k} \,,
\end{align}
\eqref{expansion} with the replacement $\nu_\ell^{(d)} \to u$, and the reflection formula of the gamma function
\begin{align}
    \Gamma (z) \Gamma (1-z) = \frac{\pi}{\sin (\pi z)}
\end{align}
with $z=u-\frac{p+1}{2}+1$, 
the coefficient of the logarithmic divergent part becomes
\begin{align}
    \frac{(-2)^{\frac{p}{2}+1}}{\Gamma(p+1)} \sum_{k=0}^{\frac{p}{2}} \beta_{k,p+2} \frac{(-1)^{k}}{2^{2k}(2k+1)} 
    = \frac{2(-2)^{\frac{p}{2}+1}}{\pi \Gamma(p+1)} \int_0^{\frac{1}{2}} \! \dd u \, \cos (\pi u) \Gamma \left( \frac{p+1}{2} +u \right) \Gamma \left( \frac{p+1}{2} -u \right) \,.
\end{align}
We find a relation
\begin{align}
    F_{\Delta_\text{D}}[\mathbb{H}^{p+1} \times \mathbb{S}^1] - F_{\Delta_\text{N}}[\mathbb{H}^{p+1} \times \mathbb{S}^1] = -2 F_\text{univ} [\mathbb{S}^p] \,,
\end{align}
where the universal free energy on $\mathbb{S}^p$ is given by \eqref{free_energy_univ}.

\paragraph{Odd $p$}

From \eqref{anomaly_even} and \eqref{finite_even} with \eqref{b.7}, the difference of the free energies is given by 
\begin{align}
\begin{aligned}
    &F_{\Delta_\text{D}}[\mathbb{H}^{p+1} \times \mathbb{S}^1] - F_{\Delta_\text{N}}[\mathbb{H}^{p+1} \times \mathbb{S}^1] \\
     &\qquad = - 2 c_{p+1} \, r_{p+1} \sum_{k=1}^{\frac{p+1}{2}} \alpha_{k,p+2} \left( f_k \left( \frac{1}{2}\right) - f_k \left( - \frac{1}{2}\right) \right) \\
    &\qquad  =  - 2 c_{p+1} \, r_{p+1} \sum_{k=1}^{\frac{p+1}{2}} \alpha_{k,p+2} (-1)^k \left( \frac{1}{2^{2k-1}(2k-1)} +  \int_{-\frac{1}{2}}^{\frac{1}{2}} \! \dd u \, u^{2k-1} \psi (u) \right) \,.
\end{aligned}
\end{align}
By using the identity
\begin{align}
    \psi (u) - \psi (-u) = \frac{1}{u} + \pi \cot (\pi u) \,,
\end{align}
\eqref{expansion} with the replacement $\nu_\ell^{(d)} \to u$ and the reflection formula of the gamma function with $z=u-\frac{p+1}{2}+1$, we find 
\begin{align}
\begin{aligned}
    F_{\Delta_\text{D}}[\mathbb{H}^{p+1} \times \mathbb{S}^1] - F_{\Delta_\text{N}}[\mathbb{H}^{p+1} \times \mathbb{S}^1] 
    & =  2 c_{p+1} r_{p+1} \int_{0}^{\frac{1}{2}} \! \dd u \, \cos (\pi u) \Gamma \left( \frac{p+1}{2} +u \right) \Gamma \left( \frac{p+1}{2} - u \right) \\
    & = -2 F_\text{univ} [\mathbb{S}^p]\,,
\end{aligned}
\end{align}
where the universal free energy on $\mathbb{S}^p$ is given by \eqref{free_energy_univ}.

Independent of the evenness of $p$, the difference of the free energies between the two boundary conditions is proportional to the universal free energy of the fermion on $\mathbb{S}^p$,
\begin{align}
    F_{\Delta_\text{D}}[\mathbb{H}^{p+1} \times \mathbb{S}^1] - F_{\Delta_\text{N}}[\mathbb{H}^{p+1} \times \mathbb{S}^1] 
    = -2 F_\text{univ} [\mathbb{S}^p] \,.
\end{align}
This fact implies that the Neumann boundary condition for $q=2$ is trivial in the sense that the defect operator saturates the unitarity bound \eqref{bound_ell} and becomes a free field.

\subsection{Evidence for defect \texorpdfstring{$C$}{C}-theorem}

We are now in a position to compare the result in section \ref{sec5.1} with our proposed conjecture in  \cite{Kobayashi:2018lil}.
The defect free energy \eqref{defect_FE} may not be invariant under the Weyl transformation, while we expect that the difference of the defect free energies is invariant.
An RG flow from the Neumann boundary condition to the Dirichlet boundary condition is triggered by a double trace deformation as is familiar in the AdS/CFT setup \cite{Witten:2001ua,Berkooz:2002ug,Gubser:2002zh,Gubser:2002vv,Hartman:2006dy,Diaz:2007an,Giombi:2013yva}, and the difference of the universal part of the defect free energies is given by 
\begin{align}
\begin{aligned}
    \tilde{D}_\text{UV} - \tilde{D}_\text{IR} & = - \sin \left( \frac{\pi p}{2}\right) \left( F_{\Delta_\text{N}} [\mathbb{H}^{p+1} \times \mathbb{S}^1 ] - F_{\Delta_\text{D}} [\mathbb{H}^{p+1} \times \mathbb{S}^1 ] \right) \\
    & = 2 \tilde{F} [\mathbb{S}^{p}]\,,
\end{aligned}
\end{align}
where 
\begin{align}
    \tilde{F} [\mathbb{S}^{p}] = \frac{2r_p}{\Gamma (p+1)} \int_{0}^{\frac{1}{2}} \! \dd u \, \cos (\pi u) \Gamma \left( \frac{p+1}{2} +u \right) \Gamma \left( \frac{p+1}{2} - u \right) 
\end{align}
is positive for any $p$.
The positivity of the sphere free energy leads to the positivity of the difference of the free energies between at UV fixed point and at IR fixed point.
In this case, our proposed defect $C$-theorem holds.

\section{Summary and Discussion}

In this paper, we studied a free Dirac fermion on $\mathbb{H}^{p+1} \times \mathbb{S}^{q-1}$ as a DCFT.
In section \ref{sec2}, we classified the allowed boundary conditions, and we found that a nontrivial boundary condition is allowed only in $q=2$.
In sections \ref{sec3} and \ref{sec4}, we computed the free energy on $\mathbb{S}^d$, $\mathbb{HS}^d$, $\mathbb{H}^d$ and $\mathbb{H}^{p+1} \times \mathbb{S}^{q-1}$ with the Dirichlet boundary condition using the zeta-function regularisation.
In particular, we obtained relations of free energies, which hold also in a conformally coupled scalar field \cite{Rodriguez-Gomez:2017kxf,Nishioka:2021uef} in section \ref{sec4.2}. In section \ref{sec5}, we computed the difference of the free energies between the Dirichlet boundary condition and the Neumann boundary condition in $q=2$ and confirmed the validity of our proposed defect $C$-theorem.

We obtained various results similar to a conformally coupled scalar case \cite{Rodriguez-Gomez:2017kxf,Nishioka:2021uef}.
However, there are several differences between the fermion case and the scalar case.
The first difference comes from the codimension of the defect which allows nontrivial boundary conditions.
Nontrivial boundary conditions in the conformally coupled scalar are allowed in $q=1,2,3,4$ \cite{Nishioka:2021uef}, but the nontrivial boundary condition in the free fermion occurs in $q=2$.
The second difference is that we rigorously derive the equivalence of the free energies between $\mathbb{HS}^d$ and $\mathbb{H}^d$ for arbitrary $d$.
However, for the conformally coupled scalar \cite{Nishioka:2021uef}, the equivalence of the free energies between $\mathbb{HS}^d$ and $\mathbb{H}^d$ is checked only numerically because a nontrivial identity among Bernoulli polynomials are required for a proof of the equivalence for arbitrary $d$.

In section \ref{sec2.2}, we gave a classification of boundary conditions in a free fermion, and this means that we constructed a concrete model of a DCFT.
A task to derive the same classification (or defect operator) of boundary conditions by using the method in \cite{Lauria:2020emq} remains.
The concrete model would be useful for a study of DCFTs.
In this paper we discussed the non-monodromy defect where the fermion does not receive any phase around the defect. For $q=2$, there exits a monodromy defect where the fermion receives a phase around the defect in addition to the non-monodromy defect.

\acknowledgments

The author thanks T.~Nishioka for the collaboration of the related work, useful discussion and various comments on the draft of this paper.
The author thanks D.~Rodriguez-Gomez and J.~G.~Russo for comments on the draft of this paper and useful discussions.
The author also thanks Y.~Abe, Y.~Okuyama and M.~Watanabe for useful communication.
The work is supported by the National Center of Theoretical Sciences (NCTS).

\appendix

\section{List of tables}
\label{app_table}

\begin{table}[ht]
    \centering
    \begin{tabularx}{\linewidth}{cYYYYY}
        \toprule
          & $\cdot$ & $\BS^1$  & $\BS^2$             & $\BS^3$       & $\BS^4$           \\ \hline
         $\cdot$ &  & $0$    & $\frac{1}{3}$  &   $0$  & $-\frac{11}{90}$   \\
         $\BH^2$ & $\frac{1}{6}$ & $0$  & $-\frac{11}{90}$   & $0$       & $\frac{191}{3780}$    \\
         \rowcolor{gray!10} $\BH^3$ & $0$ & $-\frac{37}{1440}$  & $0$               & $\frac{191}{7560}$ &   $0$  \\
         $\BH^4$ & $-\frac{11}{180}$ & $0$  & $\frac{191}{3780}$   & $0$       & $-\frac{2497}{113400}$     \\
         \rowcolor{gray!10} $\BH^5$ & $0$ & $\frac{407}{40320}$  & $0$               & $-\frac{151951}{14515200}$  &   $0$   \\
         $\BH^6$ & $\frac{191}{7560}$ & $0$   & $-\frac{2497}{113400}$   & $0$       & $\frac{14797}{1496880}$  \\
         \rowcolor{gray!10} $\BH^7$ & $0$ & $-\frac{124603}{29030400}$   &   $0$   &   $\frac{4384643}{958003200}$   &   $0$  \\
         $\BH^8$ & $-\frac{2497}{226800}$ & $0$   & $\frac{14797}{1496880}$   & $0$       & $-\frac{92427157}{20432412000}$   \\
         \rowcolor{gray!10} $\BH^9$ & $0$ & $\frac{7277933}{3832012800}$    &   $0$   &   $-\frac{43105214773}{20922789888000}$   &   $0$     \\
         $\BH^{10}$ & $\frac{14797}{2993760}$ & $0$   & $-\frac{92427157}{20432412000}$   & $0$       & $\frac{36740617}{17513496000}$ \\
         \bottomrule
    \end{tabularx}
    
    \vspace{1cm}
    
    \begin{tabularx}{\linewidth}{cYYYY}
        \toprule
          & $\BS^5$ & $\BS^6$   & $\BS^7$ & $\BS^8$  \\ \hline
         $\cdot$ & $0$  & $\frac{191}{3780}$ & $0$ & $-\frac{2497}{113400}$  \\
         $\BH^2$  &  $0$   & $-\frac{2497}{113400}$   &   $0$ & $\frac{14797}{1496880}$  \\
         \rowcolor{gray!10} $\BH^3$  &   $-\frac{33533}{2903040}$ &   $0$  &  $\frac{726491}{136857600}$ & $0$ \\
         $\BH^4$ & $0$   & $\frac{14797}{1496880}$ &   $0$ & $-\frac{92427157}{20432412000}$ \\
         \rowcolor{gray!10} $\BH^5$  &  $\frac{14797}{2993760} $  &   $0$ & $-\frac{3467627767}{1494484992000}$ & $0$ \\
         $\BH^6$   & $0$   & $-\frac{92427157}{20432412000}$ & $0$ & $\frac{36740617}{17513496000}$  \\
         \rowcolor{gray!10} $\BH^7$  &  $-\frac{4609862003}{2092278988800}$   &   $0$ &   $\frac{36740617}{35026992000}$ & $0$ \\
         $\BH^8$ & $0$   & $\frac{36740617}{17513496000}$ &   $0$ & $ -\frac{61430943169}{62523180720000}$ \\
         \rowcolor{gray!10} $\BH^9$ & $\frac{50453696437}{50214695731200}$  &   $0$ &   $-\frac{20200704144983}{41811420119040000}$ & $0$ \\
         $\BH^{10}$ & $0$   & $ -\frac{61430943169}{62523180720000}$  &   $0$ & $\frac{23133945892303}{49893498214560000}$ \\
         \bottomrule
    \end{tabularx}
    \caption{The bulk anomalies $A[\BH^{p+1}\times \BS^{q-1}]$ and the defect anomalies $\mathcal{A}[\BH^{p+1}\times \BS^{q-1}]$  (shaded) on $\BH^{p+1}\times \BS^{q-1}$ with the Dirichlet boundary condition.}
     \label{tab:F_p_q}
\end{table}

\clearpage

\begin{table}[ht]
    \centering
        \begin{tabularx}{\linewidth}{cY}
        \toprule
        $\CM$ & $F_\text{fin}[\CM]$ \\
        \hline
        $\BS^2$  &
         $ 4\, \zeta'(-1)$
        \\ 
        $\BS^3$ & $\frac{1}{4}\,\log 2+\frac{3}{8 \pi^2}\, \zeta(3)$ \\ 
        $\BS^4$ & 
         $-\frac{4}{3}\,\zeta'(-1) + \frac{4}{3}\,\zeta'(-3)$ 
        \\
        $\BS^5$ & $-\frac{3}{32}\,\log 2 -\frac{5}{32\pi ^2}\,\zeta (3)-\frac{15}{64 \pi ^4}\,\zeta(5)$\\
        $\BS^6$ & 
        $\frac{8}{15}\,\zeta'(-1) - \frac{2}{3}\,\zeta'(-3) + \frac{2}{15}\,\zeta'(-5)$
        \\
        $\BS^7$ & $\frac{5}{128}\,\log 2 + \frac{259}{3840 \pi ^2}\,\zeta(3)+\frac{35}{256 \pi^4}\,\zeta (5)+\frac{63}{512 \pi^6}\,\zeta(7)$ \\
        $\BS^8$ & 
        $-\frac{8}{35}\,\zeta'(-1) + \frac{14}{45}\,\zeta'(-3) - \frac{4}{45}\,\zeta'(-5) + \frac{2}{315}\,\zeta'(-7)$
        \\
        $\BS^9$ & $-\frac{35}{2048}\,\log 2 -\frac{3229}{107520 \pi^2}\, \zeta(3) -\frac{141}{2048 \pi ^4}\,\zeta (5) -\frac{189}{2048 \pi^6}\, \zeta(7)-\frac{255}{4096 \pi ^8}\, \zeta(9)$ \\
        $\BS^{10}$ & 
        $\frac{32}{315}\, \zeta'(-1) - \frac{82}{567}\, \zeta'(-3) + \frac{13}{270}\,\zeta'(-5) - \frac{1}{189}\,\zeta'(-7)+ \frac{1}{5670}\,\zeta'(-9)$ 
        \vspace*{0.2cm}
        \\
        \hline
        Odd $d$ & $F_\text{fin}[\mathbb{HS}^d]= \frac{1}{2}F_\text{fin}[\mathbb{S}^d]$ \\
        Even $d$ & $F_\text{fin}[\mathbb{HS}^d]= \frac{1}{2}F_\text{fin}[\mathbb{S}^d]$ 
            \vspace*{0.3cm}
            \\ \hline
        Odd $d$ & $0$ \\
        Even $d$ & $F_\text{fin}[\mathbb{H}^d]= \frac{1}{2}F_\text{fin}[\mathbb{S}^d]$ 
            \vspace*{0.3cm}
            \\ \hline
        Even $p$ & $F_\text{fin}[\BH^{p+1}\times \BS^{q-1}]=0$ \\
        Odd $p$ & $F_{\text{fin}}[\BS^{d}] = F_{\text{fin}}[\BH^{2k} \times \BS^{d-2k}]$\quad 
for $k=1,\cdots ,\ceil{d/2}-1$\\
    \bottomrule
    \end{tabularx}
    \caption{Table of the finite parts of $F_\text{fin}[\BS^d]$, $F_\text{fin}[\mathbb{HS}^d]$, $F_\text{fin}[\mathbb{H}^d]$, and $F_\text{fin}[\BH^{p+1}\times \BS^{q-1}]$ with the Dirichlet boundary condition.}
     \label{tab:Fin_p_q}
\end{table}

\section{Detail derivation of (\ref{fkm})}
\label{app2}

In this appendix, we give a detailed derivation of \eqref{fkm}.

To perform the integral 
\begin{align}
    f_k (m) = \int_0^\infty \! \dd \omega \,  \frac{\omega^{2k-1}}{\mathrm{e}^{2\pi \omega}-1} \log (\omega^2 + m^2 ) \,,
\end{align}
we first take the derivative respect to $m$,
\begin{align}
    \partial_m f_k (m) &= 2m g_k (m) \,, \\
    g_k (m) & = 
    \int_0^\infty \! \dd \omega \,  \frac{\omega^{2k-1}}{(\mathrm{e}^{2\pi \omega}-1)(\omega^2 + m^2)} \,.
\end{align}
Since $g_k (m)$ satisfies the recursion relation
\begin{align}
    g_{k+1} (m) = - m^2 g_k (m) + (-1)^{k+1} \frac{B_{2k}}{4k} \,,
\end{align}
with the initial condition
\begin{align}
    g_1 (m) = \frac{1}{2} \left( \log m -\frac{1}{2m} -\psi (m) \right) \,,
\end{align}
the solution is given by
\begin{align}
    g_k (m) = (-1)^{k-1} m^{2k-2} \left( g_1 (m) - \sum_{l=1}^{k-1} m^{-2l}  \frac{B_{2l}}{4l} \right) \,.
\end{align}
By integrating $2m g_k (m)$ from $0$ to $m$, 
$f_k (m)$ can be evaluated as
\begin{align} \label{b.7}
\begin{aligned}
    f_k(m) & = (-1)^k \left[ - \zeta'(1-2k) + \frac{m^{2k-1}}{2(2k-1)} + \frac{m^{2k}}{4k}  \left( \frac{1}{k} - \log( m^2)  \right) \right.  \\
    & \qquad \qquad \left. + \int_0^m \! \dd \mu \, \mu^{2k-1} \psi (\mu) + \sum_{l=1}^{k-1} \frac{B_{2l}}{4l} \frac{m^{2k-2l}}{k-l} \right] \,,
\end{aligned}
\end{align}
where we use 
\begin{align}
\begin{aligned}
     f_k (0) & = 2 \int_0^\infty \! \dd \omega \frac{\omega^{2k-1}}{\mathrm{e}^{2\pi \omega}-1} \log \omega \\
     &= (-1)^{k-1} \zeta'(1-2k) \,.
\end{aligned}
\end{align}
The remaining integral is performed using a formula in
\cite{adamchik1998polygamma,espinosa2001some}
\begin{align}
\begin{aligned}
     \int_0^m \! \dd \mu \, \mu^{n} \psi (\mu) & = (-1)^n \left( \frac{B_{n+1}H_n}{n+1} - \zeta'(-n) \right) \\
     & \qquad + \sum_{r=0}^{n} (-1)^r \binom{n}{r} m^{n-r} \left( \zeta'(-r,m) -  \frac{B_{r+1}(m) H_r}{r+1} \right) \,.
\end{aligned}
\end{align}

\bibliographystyle{JHEP}
\bibliography{Defect}

\end{document}